# Information transfer by entangled photons without auxiliary non-quantum channel

*Levente Szabó[1], Pál Andor Maák[2]*

[1]Department of Sociology and Communication, Budapest Univ. of Technology and Economics, Hungary,
szabo.levente@gtk.bme.hu

[2]Department of Atomic Physics, Inst. of Physics, Budapest Univ. of Technology and Economics, Hungary,
maak@eik.bme.hu

## Abstract

In this paper we present a theoretical analysis of the faster than light communication possibility based on entangled photons. We analyze designs that may be capable to solve the problem of direct information transfer between members of an entangled photon pairs. We consider that experimental verifications can confirm or even refute this. Our hypothesis was that most proofs of the no-communication theorem are based on a certain set of conditions, and it is possible to provide a broader set of conditions that allow the establishment of entangled states as quantum information channels, without using a classical channel. One basic unit of the proposed design transforms the polarization state of one member of an entangled photon pair into a spatial superposition state. Thus, after the polarization measurement performed on one member, which eliminates the entanglement, the quantum information is maintained in the spatial superposition state of the other member. This can be recovered by a particular measurement based on spatial interference. We have shown that solutions with so-called symmetric functions lead to average results that corresponds to the no-communication theorem. However, using asymmetric functions the averaged measurement results calculated in a prescribed time window can distinguish the types of measurements performed on the other member of the pair. This can establish a communication code that enables faster-than-light information sharing under specific conditions. There may be also further theoretical consequences: a significant extension of the quantum mechanical nonlocality principle.

## 1. Introduction

The impossibility of faster-than-light information transfer is not self-evident in quantum mechanics. [1] [2] This explicit limit stems from the special relativity theory. [3] Quantum mechanics, as well as special (and general) relativity theory, offer descriptions of the physical world based on different foundations. Therefore, reconciling or linking these two theoretical descriptions has been a priority in physics from the beginning. [4] [5] The relativistic quantum mechanics and quantum field theories created in this way include the speed-of-light constraint on the propagation of effects [6], thus meeting the expected causal relationships in special relativity theory [7].

The feature of non-locality is demonstrated by the so-called EPR paradox, which describes two entangled particles that can be spatially separated but belong to the same wavefunction. When a measurement is made on one particle, the entire wavefunction collapses in a specific way, and the other particle ends up in a corresponding state. This results in correlated measurement outcomes for the two particles (completely in the case of pure Bell states). (For a brief description and formal basic criteria see Appendix A1, for an overview of the problem of nonlocality and entanglement see Appendix A2.)

The question of paramount importance is why, if measurements can determine each other, this cannot be used for instant information transfer. The no-communication theorem provides a kind of answer to this. [3] According to this theorem, in the case of entangled quantum states, information that cannot be transmitted by the measurement of one subsystem to the other subsystem, so that it appears in its measurement results – although there is a correlation between the measurements.

We question the widely accepted theorem that the nonlocality offered by the entanglement of quantum particles does not allow information transfer without auxiliary classical information channels. We argue that the conditions defined for various versions of the no-communication theorem are not general. We define a possible broader set of conditions under which a processing method may allow direct information transfer between parties. The task of such a construction is to properly manipulate the quantum information carried by entangled pairs of photons (in the form of wave functions) and allow measurements on the receiver side with outcomes depending on the action on the sender side. We demonstrate that it is possible to provide wave functions that locally contain information about the entire systems. We show that this forms the basis of the communication possibility solely via entangled photons. The main novelty is in the method used to extract information from the wave functions originally hidden in the joint quantum state, and to make the direct relation between the action of the sender and the measurement result. The proposed method is based on a statistical evaluation of averaged photon detector signals, the temporal period used on the sender and receiver side must be agreed in advance but synchronization is not required.

The presented train of thought is entirely theoretical, we demonstrate through simple mathematical derivations the premise, our goal was to apply the simplest formalism possible. The formulas apply solely to quantum mechanical systems containing only two photons, and we do not provide the more general relationships to which these are partial cases. Our reasoning is logically based on the premise that if something can be proven in a specific case, then the general conclusion that communication with entangled states is possible is true. In the presented context case, we only rely on pure Bell states, a mixture of different states, as characteristic to real life sources, is not considered. The task is to demonstrate the idea, practical realizations are left for future research.

**1.1 Beyond the no-communication theorem**

There are several different types of proofs of the no-communication theorem, but fundamentally each one is related to the Bell theorem. [3] [7] [8] [9] [10] The Bell theorem states that if the physical properties targeted by the measurements are not determined beforehand and locally, then the measurements will be random from a local perspective for each entity. In the no-communication theorem, the local perspective is if we only consider the measurement results of one of the entangled particles. Therefore, if, as previously stated, there is a correlation between the measurement results of two entangled particles, but at the same time, from a local point of view, the measurement results of one particle are random, then from a local point of view, the measurement results of the other particle will also be random. However, the randomness seen in the second particle from a local perspective would be the same as if no measurement had occurred on the first particle (or as if the first measurement had occurred on the second particle). Therefore, the conclusion of the no-communication theorem is that measurements of entangled quantum states are correlated, but information about this correlation cannot be obtained from local measurements, and thus the information cannot be transmitted through the correlation. The "global" information about the correlation can only be established through a classical channel limited by the speed of light.

Thus, the no-communication theorem offers an answer to a contradiction in relation to entanglement phenomena. The contradiction lies between the non-locality of quantum mechanics and the principle of locality of special relativity theory or field and space theories based on it. In the former case, immediate correlations, in some interpretations immediate actions, are created, while the latter restrict the propagation of effects through localities at the speed of light. The relationship between these two types of theories is not yet fully clarified today – just as the scope of validity and degree of generalizability of the Bell theorem is a subject of research. The fundamental question is whether theories representing local realism can accept and interpret entanglement phenomena, so it is also a question whether special relativity theory can be a framework theory in which entanglement phenomena can be described (for example, in reference frames).

However, the conclusion derived from the no-communication theorem can only offer a partial solution. On the one hand, it proves that the information content of measurements of entanglement phenomena can only be conveyed at the speed of light, which satisfies the requirement of special relativity theory [11]. On the other hand, the no-communication theorem evaluates only one aspect of the correlation arising from measurements of entanglement phenomena, namely the informational aspect. Because the correlation can be immediate, but the propagation of information regarding this is of finite speed.

In contrast, in relativity theory, the "correlations" – such as simultaneity of events – and the information regarding them coincide. Here, the "correlations" have meaning precisely in terms of how they appear as "correlations" to observers, for example, as light signals arriving simultaneously in a given reference frame. Therefore, the "correlation" and the information about it are not separated – unlike what is seen in quantum mechanics. Similarly, in relativity theory, for photons, the effect and the information about the effect coincide: the observer is informed by the effect itself. In the case of entanglement phenomena, the measurement of one subsystem has an immediate effect on the other subsystem (if we accept this interpretation), but the information about the occurrence of the effect is separate from this. Therefore, the no-communication theorem makes only the informational aspect of the correlation of measurements related to entanglement phenomena acceptable to relativity theory. However, this "informational" model may still hold many developments.

## 2. A broader concept of measurement conditions

The conclusion of the no-communication theorem is that „faster than light" information transfer is generally not possible, there may be no circumstances to allow it. On the other hand our opinion is that if the conditions under which the demonstration of the theorem had place are not general the conclusion drawn for the theorem is also not general.

Some basic conditions of the no-communication theorem are within the definition of the measurement concept. Basic characteristics of the measurement concept used in the demonstration of the theorem:

1. the measurements is targeting one particular physical property of the subject;

2. the result of the measurement is a definite value;

3. the measurement on an entangled pair cannot be repeated, it is non-recurring.

We emphasize that the theorem reduces the possibility of information transfer to measurements with particular and definite outcomes based assuming the above listed characteristics of the measurement.

We think that these restrict the validity of the theorem and think that expanding the characteristics of the measurement concept leads to communication possibilities outside its validity range.

In the next chapters we analyze a measurement concept realized in a measurement assembly that extends the above characteristics of measurement.

1. The measurement is not intended to measure only one parameter of the photon, the measuring instrument performs simultaneously the measurement of polarization and photon path

2. The measurements does not target a single quantum-state, the polarization and path together represent a superposition of both quantum states

3. The outcome of the measurement is not a strictly determined value, the measurements are only weak, which did not result in a complete reduction of the wave function, the quantum states persist after the weak measurement

4. The result of the measurement is not unique on an entangled pair, after the first weak measurement further measurements are allowed by the persisting quantum state, who may be independent on the result of the first

This extended concept of measurement means that in certain combinations the collapse of the wave function due to the measurement is not full, the measurement result is not unique and has an amount of uncertainty, then this will result in the persistence of entanglement. We expect also that if the measurement concept is handling two quantum properties, which are related to each other, measurement of one property does not lead to complete loss of information since it is further carried by the other. In our proposed concept the polarization state is related to the photon path: while the superposed state of the polarization is collapsed the spatial position of the photon remains in a superposed state related to that of the polarization anterior to the measurement event. The superposed spatial position state can be further measured and evaluated. With this transfer of the information between polarization and spatial position states we expect that the conclusion of the no-communication theorem, which assumes that one measurement erases all information, is not generally valid.

Particularity of this concept based on „weak" measurements is that the final measurement result is depending on both the initial state of the photon and changes in their states during the transfer through the arrangement. Our proposed concept is based on statistical evaluation of detector signals, at least information transfer in our system requires measurement of multiple photons within a time window. Hence this information transfer can not be considered as instantaneous, in the sense that detection of a photon immediately results in an information (e.g. bit) content on the other side. However, the distance between the partners may induce a higher time needed for light propagation between them than that needed for the measurement evaluation.

## 3. Conceptual description of the construction

Below we present a device that meets the requirements described in Chapter 2. Fig. 1 is a conceptual representation, based on that we describe the quantum mechanical and quantum informatics processes and measurement-detection results.

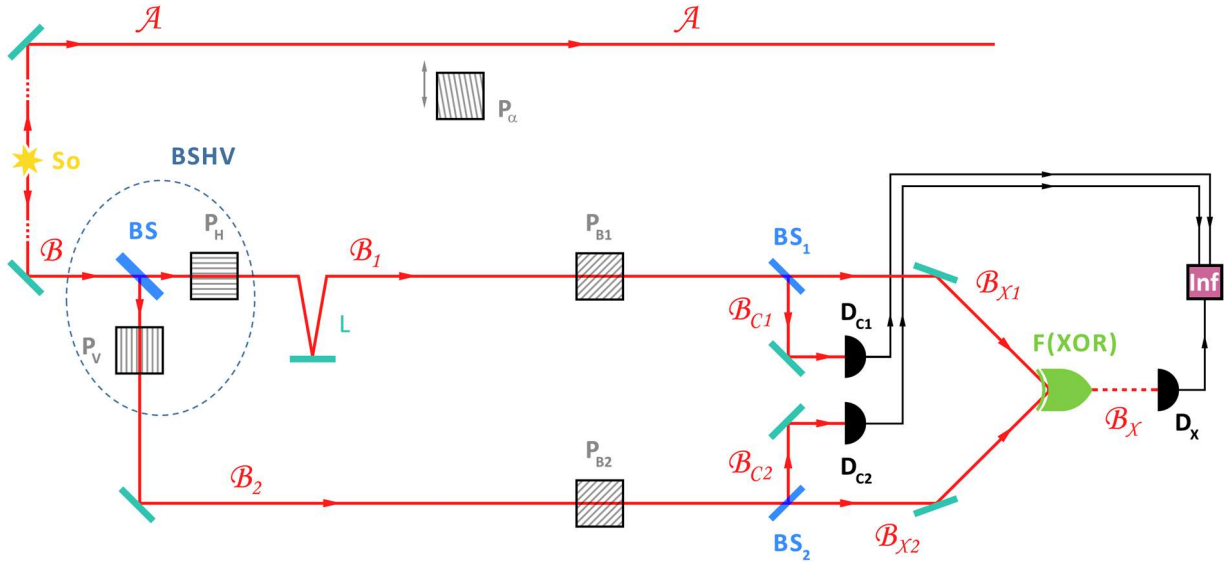

**Fig. 1. Schematic of the principle construction with an unspecified F(XOR) gate.** So – photon source; A – Alice branch; B – Bob branch; $B_1$, $B_2$ – Bob's split branches; BS – beam splitter; $P_H$ – horizontal polarizer; $P_V$ – vertical polarizer; BSHV – the unit consisting of BS, $P_H$, $P_V$ elements; L – distance regulator; $P_{B1}$, $P_{B2}$ – 45° polarizer on branch $B_1$, $BS_1$, $BS_2$ – beam splitters; $B_{C1}$, $B_{C2}$ – control branches; $D_{C1}$, $D_{C2}$ – control detector system; F(XOR) – special XOR gate; $B_{X1}$, $B_{X2}$ – input branches of the F(XOR) gate; $B_X$, – output branch of the F(XOR) gate; $D_x$ – detector (system) after F(XOR) gate; Inf – data processing unit; $P_\alpha$ – insertable-removable α angle polarizer on the Alice branch

So is a sufficiently weak photon source, emitting entangled photons in two directions, on branches A (Alice) and B (Bob). In principle, a sufficient time interval is provided between the exits of individual photon pairs so that they are separated from each other and do not overlap each other in the subsequent processes.

On branch B, the photon arrives at the BS beam splitter, which splits this path into sub-branchs $B_1$ and $B_2$. This allows the quantum state of the photon to be a spatial superposition. At the same time, the PH and PV polarizers allow different polarizations on the two paths, only horizontal (H) and vertical (V) polarization values. Although the sequence of the BS beam splitter and $P_H$, $P_V$ polarizers in the schematic diagram indicates separate operations, it is advisable to consider these elements as a system (denoted as BSHV) and to interpret the spatial division and polarization value assignment as one operation from the point of view of the calculations (this requirement it can also be done in the case of a technical implementation, e.g. with a polarizing beamsplitter). BSHV opens up the perspective that manipulations on one of the coupled spatial and polarization superposition quantum states will result in "traces" on the other state as well, from which information may be obtained.

The function of the distance regulator L is to ensure that the photon, after entering the BS beam splitter, travels the same distance along paths B1 and B2 to the inclined polarizers $P_{B1}$, $P_{B2}$.

The polarizers $P_{B1}$, $P_{B2}$ oriented diagonally provide parallel polarized photons from orthogonally polarized ones, when pass through. $BS_1$ and $BS_2$ beam splitters split the $B_1$ and $B_2$ branches. Photons can propagate with equal probability on branches $B_{C1}$ and $B_{C2}$, and on branches $B_{X1}$ and $B_{X2}$. In the latter case, the components of the photons enter a gate with an F(XOR) function, the results of which are recorded by the $D_x$ detector. The results of the $D_{C1}$ and $D_{C2}$ detectors perform a kind of "control" function, their signal is used for reference and normalization.

We do not go into detail here about the F(XOR) gate. We will first give a general description of this later, and then we will also develop different specific solutions. (The dotted line representation of the $B_X$ branch indicates that the effect of the F(XOR) gate is not specified here. $D_X$ may also represent a detector system.)

Statistical data is aggregated and processed in the Inf unit, if a time window can be defined, within which a certain amount of photons flows through the entire system (but separated from each other in time).

By default, the photon entering branch A from the photon source So is not subject to any manipulation. It is possible to create a case where a polarizer $P_\alpha$ with an angle α is placed on branch A. The question will obviously be whether the result of the Inf data processing unit of branch B shows a difference between the case without manipulation and this case? Just as the question will be whether different applications of $P_\alpha$ polarizer with angle α lead to different results on branch B?

## 3.1 Mathematical characteristics of main elements

### 3.1.1 The BSHV for entangled photons

The BSHV functions as a weak measuring device. a.) The measurement means that, after the polarizers, the state of the photon (component) can only be H-polarized on the $B_1$ track, and V-polarized on the $B_2$ track. Whatever the state of the photon before entering the unit. b.) On the other hand, however, a photon with an arbitrary polarization state will not necessarily travel along a $B_1$ or $B_2$ path with a probability of unity. It can be indeterminate in the sense that the two paths will have superposition components. Conditions a.) and b.) together represent the concept of weak measurement.

We can conceive that passing through the BSHV performs a weak measurement, which does not change the form of the full state vector of the full system:

$$\frac{1}{\sqrt{2}}(|a_1\rangle|b_1\rangle + |a_2\rangle|b_2\rangle) \implies l\frac{1}{\sqrt{2}}(|\uparrow\rangle_A|\uparrow\rangle_{B2} + |\rightarrow\rangle_A|\rightarrow\rangle_{B1}) \qquad (1)$$

This is based on the transformation, which is well explored in the literature, and its detailed description and derivation is provided in Appendix A3.

The coefficient *l* adds the losses associated to the passing through the BSHV. Using appropriate technical solutions this loss can be minimized. Since the outcome of the system is found in the ratio of the detector signals $D_{C1}$, $D_{C2}$ and $D_X$, this coefficient will vanish, and must not be taken into account.

While the form of the state vector describing the photon pair remains unchanged after that one member of the pair passed through the BSHV, we can conclude that the entanglement remained between the members. This is one merit of the „weak measurement" performed by this beam splitter.

The functionality of the BSHV leads to another important result, which is also observed in the formula: when the possible polarization states of the photon in the B side are separated spatially, then the related phases are also representing this spatial division. We think that this spatial separation makes the relative phase measurable, allows its manipulation and we can gather at least information about this specific parameter of the full state. Our vision is that this spatial separation can provide the possibility of communication with entangled photons without the need for correlation information.

### 3.1.2 F(XOR) gate

The particular task of the F(XOR) gate is to transfer the relation between $B_{X1}$ and $B_{X2}$ branches to information on a detector. The information belonging to the two branches is generated by the weak measurement performed by the BSHV. $B_1$ branch contains only horizontally polarized, $B_2$ branch contains only vertically polarized components, which are unified by the oblique parallel polarizers before the F(XOR) gate.

The implementation of a XOR gate usually contains means that produces interaction between two beams (photons) having particular values generating new outcome depending on the inter-relation. Using a binary coding the truth-table of the XOR gate:

| $B_{X1}$ | $B_{X2}$ | $B_{X1}$ XOR $B_{X2}$ |
|---|---|---|
| 0 | 0 | 0 |
| 0 | 1 | 1 |
| 1 | 0 | 1 |
| 1 | 1 | 0 |

Table 1. Truth-table of the XOR gate

One possible implementation is a spatially or temporally filtered interference, which according to the 4th row may produce zero output even if both inputs are nonzero. The XOR gate can also be used for scaling, like expressed in the following 4th row:

| $B_{X1}$ | $B_{X2}$ | $B_{X1}$ S(XOR) $B_{X2}$ |
|---|---|---|
| 1 | 1 | $0 \leq S < 1$ |

Table 2. S(XOR): scalable XOR gate

In our case we want to exhaust this latter function to measure the relation between the probability amplitude of branches $B_{X1}$ and $B_{X2}$. In general case the wave-function components of the $B_{X1}$ and $B_{X2}$ branches written in a common orthonormal basis, corresponding to the arbitrary polarization state of the :

$$|\psi\rangle = f_{BX1}(\alpha) \cdot |\psi\rangle_{BX1} + f_{BX2}(\alpha) \cdot |\psi\rangle_{BX2} \qquad (2)$$

where

$|\psi\rangle_{BX1}$, $|\psi\rangle_{BX2}$ – are the components of the photon's $|\psi\rangle$ wave function belonging to branches $B_{X1}$ and $B_{X2}$, respectively

$\alpha$ – the angle of a unit vector closed with the x axis used for defining the spatial superposition state of the photon propagating in branches $B_{X1}$ and $B_{X2}$.

$f_{BX1}$, $f_{BX2}$ – the probability amplitudes of the function components belonging to the $B_{X1}$ and $B_{X2}$ branches, respectively

When we interpret the probabilities calculated from the probability amplitudes as photon number ratios, we will get a relation F(XOR) between the photon number available in branch $B_X$ at the output of the gate:

| branch | $B_{X1}$ | $B_{X2}$ | $B_X \equiv B_{X1}$ F(XOR) $B_{X2}$ |
|---|---|---|---|
| photon count ratio | $\dfrac{N_{BX1}}{N}$ | $\dfrac{N_{BX2}}{N}$ | $\dfrac{N_{BX}}{N}$ |
| $\alpha$ | $[f_{BX1}(\alpha)]^2$ | $[f_{BX2}(\alpha)]^2$ | $[f_X(\alpha)]^2 = \{f_X[f_{BX1}(\alpha), f_{BX2}(\alpha)]\}^2$ |

Table 3. General definition of the F(XOR) gate

where:

$f_X$ – is the function representing the interaction outcome of the probability amplitudes $f_{BX1}$ and $f_{BX2}$ valid for branches $B_{X1}$ and $B_{X2}$, respectively;

$N_{BX1}$, $N_{BX2}$, $N_{BX}$, – the number of photons in branches $B_{X1}$, $B_{X2}$ and $B_X$, respectively;

N – the total number of photons entering the F(XOR) gate (N = $N_{BX1}$ + $N_{BX2}$);

The definition of the F(XOR) allows to generate all rows of the truth-table, when setting $[f_{BX1}(\alpha)]^2$ = 0 or 1 and $[f_{BX}(\alpha)]^2$ = 0 or 1. The $f_X$ function implements scaling according to angle α, hence it plays a definitive role in the operation of the construction.

The task of the further deductions is to determine the probability amplitudes $f_{BX1}$, $f_{BX2}$ and $f_X$ for different circumstances occurring during the joint activity of the two sides A and B. We can foresee right now that amplitudes $f_{BX1}$, $f_{BX2}$ follow immediately from the definition of the total construction and mathematical description of the quantum-mechanical processes taking place within it. The resulting amplitude $f_X$ max be of different value depending on the applied implementation.

## 4. Quantum informatics processes of construction

In the following, we describe two cases that show the passage of photons of different states through the theoretical setup given in Fig. 1. In the first case, an entangled photon pair passes through branches A and B, and the photon in branch A is not subjected to any operation (case I.), in the second case, an entangled photon pair also passes through branches A and B, and a measurement is performed on the photon in branch A (case II. ). In branch B, the processing of the results of a detector system creates information about the processes that have passed through the structure. In the first step, the F(XOR) gate is represented by a phase-shifting beam splitter. Other solutions will be presented later.

### *4.1 Entangled pair of photons (case I.)*

In the way of the photon in the B side there are polarizers oriented at 45° relative to the local horizontal-vertical coordinate system in both $B_1$ and $B_2$ branches. To represent the propagation through these polarizers, the photon states should be decomposed to polarizations parallel and perpendicular to this tilted polarizer orientation. Further question is that this new decomposition is also conserving the invariance of the mathematical description of the state to polarization orientations as is the case for simple polarization entangled pairs?

Transscribing the vectors in the new polarization base rotated by 45°:

$$\frac{1}{\sqrt{2}}(|\uparrow\rangle_A|\uparrow\rangle_{B2} + |\rightarrow\rangle_A|\rightarrow\rangle_{B1}) =$$

$$= \frac{1}{\sqrt{2}}\Big[\frac{1}{\sqrt{2}}(|\nwarrow\rangle_A + |\nearrow\rangle_A)\frac{1}{\sqrt{2}}(|\nwarrow\rangle_{B2} + |\nearrow\rangle_{B2})$$

$$+ \frac{1}{\sqrt{2}}(-|\nwarrow\rangle_A + |\nearrow\rangle_A)\frac{1}{\sqrt{2}}(-|\nwarrow\rangle_{B1} + |\nearrow\rangle_{B1})\Big] =$$

$$= \frac{1}{2\sqrt{2}}(|\nwarrow\rangle_A|\nwarrow\rangle_{B2} + |\nwarrow\rangle_A|\nearrow\rangle_{B2} + |\nearrow\rangle_A|\nwarrow\rangle_{B2} + |\nearrow\rangle_A|\nearrow\rangle_{B2}$$

$$+ |\nwarrow\rangle_A|\nwarrow\rangle_{B1} - |\nwarrow\rangle_A|\nearrow\rangle_{B1} - |\nearrow\rangle_A|\nwarrow\rangle_{B1} + |\nearrow\rangle_A|\nearrow\rangle_{B1})$$

In this rotated superposition the polarization direction invariance is not more conserved because of the spatial separation of the components. The superposed states do not represent a pure Bell state. We would get the invariance back, if we would remove the spatial separation and would join the two branches without any further manipulation, assuring that the phases accumulated during propagation in the two branches are perfectly equal. This also means that the original entanglement and corresponding joint phase of the pair are maintained after the spatial separation.

Only the terms containing polarization components parallel to the tilted polarizer orientation will represent states after passing through the polarizer, the pairs containing orthogonally polarized photons disappear: the state representing the photons, which passed through the polarizers must be normalized.

$$\Rightarrow \quad \tfrac{1}{2}(|\nwarrow\rangle_A|\nearrow\rangle_{B2} + |\nearrow\rangle_A|\nearrow\rangle_{B2} - |\nwarrow\rangle_A|\nearrow\rangle_{B1} + |\nearrow\rangle_A|\nearrow\rangle_{B1}) \qquad (3)$$

This form does not fulfill the entanglement criterion expressed in Equation (a3): according to Equation (a2) we will get from the coefficients the formal relation ru=st. This also indicates that after the tilted polarizers entanglement is broken and the members of the pair will be independent.

After the tilted polarizers beam splitters $BS_1$ and $BS_2$ will split the branches $B_1$ and $B_2$ with equal probabilities. The superposition state of the photons after division:

$$\Rightarrow \quad B_{C1} + B_{C2} + B_{X1} + B_{X2}: \tfrac{1}{2\sqrt{2}}(|\nwarrow\rangle_A|\nearrow\rangle_{BC2} + |\nearrow\rangle_A|\nearrow\rangle_{BC} - |\nwarrow\rangle_A|\nearrow\rangle_{BC} + |\nearrow\rangle_A|\nearrow\rangle_{BC1})$$

$$+ \tfrac{1}{2\sqrt{2}}(|\nwarrow\rangle_A|\nearrow\rangle_{BX2} + |\nearrow\rangle_A|\nearrow\rangle_{BX2} - |\nwarrow\rangle_A|\nearrow\rangle_{BX1} + |\nearrow\rangle_A|\nearrow\rangle_{BX1})$$

The terms arriving to detector $D_C$ are, separated:

$$B_{C1}: \tfrac{1}{2\sqrt{2}}(-|\nwarrow\rangle_A|\nearrow\rangle_{BC1} + |\nearrow\rangle_A|\nearrow\rangle_{BC}) = f_{BX1}(\alpha) \cdot |\psi\rangle_{BX} \qquad (4)$$

$$B_{C2}: \tfrac{1}{2\sqrt{2}}(|\nwarrow\rangle_A|\nearrow\rangle_{BC2} + |\nearrow\rangle_A|\nearrow\rangle_{BC}) = f_{BX2}(\alpha) \cdot |\psi\rangle_{BX} \qquad (5)$$

With this, we determined functions $f_{BX1}$ and $f_{BX2}$ (see Equation (2)), which are measured on branches $B_{C1}$ and $B_{C2}$. Their characteristic is that they are independent of α and the members of the B system are connected to the possible values of the A system by a tensor product.

These are measured independently, the detection probability density will become:

$$D_C = D_{C1} + D_{C2} = 2 \cdot \left(\tfrac{1}{2\sqrt{2}}\right)^2 + 2 \cdot \left(\tfrac{1}{2\sqrt{2}}\right)^2 = \tfrac{1}{2}$$

The branches $B_{X1}$ and $B_{X2}$ converge in the F(XOR) gate, which performs the interfering function. For the analysis, we take the simplest case, when the interference is performed by one of the outputs of a π phase shift beam splitter. Then the output function of $B_X$ branch:

$$B_X: f_X(\alpha) \cdot |\psi\rangle_{BX} = \tfrac{1}{4}(|\nwarrow\rangle_A|\nearrow\rangle_{BX2} + |\nearrow\rangle_A|\nearrow\rangle_{BX2} + |\nwarrow\rangle_A|\nearrow\rangle_{BX} - |\nearrow\rangle_A|\nearrow\rangle_{BX1})$$

$$= \tfrac{1}{4}(|\nwarrow\rangle_A|\nearrow\rangle_{BX} + |\nwarrow\rangle_A|\nearrow\rangle_{BX}) \qquad (6)$$

The other two terms fall out because of the phase shift, the sum of the remaining members gives the detection probability. The probability measured by $D_X$:

$$D_X = \left(\frac{2}{4}\right)^2 = \frac{1}{4}$$

When we repeat the measurement on high number of individual, originally entangled photon pairs independent from each other we will get as the output of the data processing unit Inf:

$$N_S = \frac{D_X}{D_C} = \frac{1}{2} \qquad (7)$$

$N_s$ is a dimensionless quantity not depending on the number of the analyzed photons, it is providing information about the ratio of the photons making successful interference in the interferometric F(XOR) gate. Its value exactly shows that only half of the total number of photons, counted by $D_C$ is providing negative interference at $D_X$. It is most surprising that this ratio is constant, but is was still expected since Equation (3) shows symmetric relation between the components of the states split to the $B_1$ and $B_2$ branches. This finally also means that if orthogonally polarized entangled photons in pure Bell superposition state enter the system and the photon propagating in part A is left intact than the output of the Inf unit, $N_s$ will be always constant, independent of the original orientation of the polarizations. This is very different from the case discussed in the next chapter, where the output produced by a non-entangled photon is heavily dependent on the orientation of the polarization relative to the local coordinate basis α.

## *4.2 Entangled pair of photons, measurement performed on branch A (case II.)*

In this case a measurement of the polarization is performed in the A side with an arbitrarily oriented polarizer, which is symbolized with a polarizer oriented at angle α relative to the local vertical axis of the side B. Then the state of the photon in side A must be decomposed according to the direction of the polarizer. We use a special notation to make this decomposition, where $|\uparrow\rangle_{\alpha A}$ denotes the state with the polarization parallel to the axis of the polarizer, and $|\rightarrow\rangle_{\alpha A}$ denotes the state with the polarization perpendicular to it. Using this we get for the previous states:

$$|\uparrow\rangle_A = |\uparrow\rangle_{\alpha A} \cdot \cos\alpha - |\rightarrow\rangle_{\alpha A} \cdot \sin\alpha$$

$$|\rightarrow\rangle_A = |\uparrow\rangle_{\alpha A} \cdot \sin\alpha + |\rightarrow\rangle_{\alpha A} \cdot \cos\alpha$$

Using this notation the entangled state before the measurement becomes:

$$\frac{1}{\sqrt{2}}\left((|\uparrow\rangle_{\alpha A} \cdot \cos\alpha - |\rightarrow\rangle_{\alpha A} \cdot \sin\alpha)|\uparrow\rangle_{B2} + (|\uparrow\rangle_{\alpha A} \cdot \sin\alpha + |\rightarrow\rangle_{\alpha A} \cdot \cos\alpha)|\rightarrow\rangle_{B1}\right)$$

In fact, the use of the polarizer in A side is equivalent to a measurement, which causes the disappearance of entanglement. The system will have two possible outcomes, depending on whether the photon in A side is transmitted or absorbed.

1) **Case II.a**, when the photon is transmitted by the A side polarizer, than it remains in the state $|\uparrow\rangle_{\alpha A}$ and the joint state with side B will be, normalized:

$$\Rightarrow \quad |\uparrow\rangle_{\alpha A} \cdot \cos\alpha \cdot |\uparrow\rangle_{B2} + |\uparrow\rangle_{\alpha A} \cdot \sin\alpha \cdot |\rightarrow\rangle_{B1} \qquad (8)$$

To prepare for meeting the oblique polarizers, polarizations in branches $B_1$ and $B_2$ should be decomposed into combination of states polarized at +/- 45°.:

$$|\uparrow\rangle_{\alpha A} \cdot \cos\alpha \cdot \frac{1}{\sqrt{2}}(|\nwarrow\rangle_{B2} + |\nearrow\rangle_{B2}) + |\uparrow\rangle_{\alpha A} \cdot \sin\alpha \cdot \frac{1}{\sqrt{2}}(-|\nwarrow\rangle_{B1} + |\nearrow\rangle_{B1})$$

After passing through the polarizers tilted at 45° only the terms containing states polarized under 45° remain, the new state normalized:

$$\Rightarrow \; |\uparrow\rangle_{\alpha A} \cdot \cos\alpha \cdot |\nearrow\rangle_{B2} + |\uparrow\rangle_{\alpha A} \cdot \sin\alpha \cdot |\nearrow\rangle_{B1} \qquad (9)$$

These are divided by the BS$_1$, BS$_2$ beam splitters into equal parts, hence we have in the paths B$_{C1}$, B$_{C2}$:

$$B_{C1}: \frac{1}{\sqrt{2}} |\uparrow\rangle_{\alpha A} \cdot \sin\alpha \cdot |\nearrow\rangle_{BC1} = f_{BX1}(\alpha) \cdot |\psi\rangle_{BX1} \qquad (10)$$

$$B_{C2}: \frac{1}{\sqrt{2}} |\uparrow\rangle_{\alpha A} \cdot \cos\alpha \cdot |\nearrow\rangle_{BC2} = f_{BX2}(\alpha) \cdot |\psi\rangle_{BX2} \qquad (11)$$

With this, we defined functions f$_{BX1}$ and f$_{BX2}$, which are measured on branches B$_{C1}$ and B$_{C2}$. Their characteristic is that they are a function of the possible values of α.

From here the signal of detector D$_C$ will be:

$$D_C = D_{C1} + D_{C2} = \frac{1}{2}(\cos^2\alpha + \sin^2\alpha) = \frac{1}{2}$$

The branches B$_{X1}$ and B$_{X2}$ converge in the F(XOR) gate performing the interference function, where interference is performed by one of the outputs of a π phase shift beam splitter. Then the output function of B$_X$ branch:

$$B_X: f_X(\alpha) \cdot |\psi\rangle_{BX} = \frac{1}{2} |\uparrow\rangle_{\alpha A} (\cos\alpha \cdot |\nearrow\rangle_{BX} - \sin\alpha \cdot |\nearrow\rangle_{BX}) \qquad (12)$$

The detector signal will be:

$$D_X = \frac{1}{4}(\cos\alpha - \sin\alpha)^2 = \frac{1}{4}(1 - \sin 2\alpha)$$

When considering high number of photons and interpreting the probability amplitudes as numbers of the detected photons, we get as the output of unit Inf:

$$N_S = \frac{D_X}{D_C} = \frac{1}{2}(1 - \sin 2\alpha) \qquad (13)$$

2.) **Case II.b** The measurement in A side may have the second outcome when the photon in her side is absorbed. In that case its polarization becomes perpendicular to the polarizer oriented at angle α relative to the vertical axis. This imposes the following state in the B side, normalized:

$$\Rightarrow \; -\sin\alpha \cdot |\uparrow\rangle_{B2} + \cos\alpha \cdot |\rightarrow\rangle_{B1} \qquad (14)$$

Rewritten using the decomposition to states polarized at +/- 45°:

$$-\sin\alpha \cdot \frac{1}{\sqrt{2}}(|\nwarrow\rangle_{B2} + |\nearrow\rangle_{B2}) + \cos\alpha \cdot \frac{1}{\sqrt{2}}(-|\nwarrow\rangle_{B1} + |\nearrow\rangle_{B1})$$

After passing through the polarizers oriented at +45°, normalized:

$$\Rightarrow \; -\sin\alpha \cdot |\nearrow\rangle_{B2} + \cos\alpha \cdot |\nearrow\rangle_{B1} \qquad (15)$$

After division on the beam splitter the photon remains in superposition state in the paths B$_{C1}$, B$_{C2}$:

$$B_{C1}: \frac{1}{\sqrt{2}} \cos\alpha \cdot |\rightarrow\rangle_{B1} = f_{BX1}(\alpha) \cdot |\psi\rangle_{BX1} \qquad (16)$$

$$B_{C2}: -\frac{1}{\sqrt{2}} \sin\alpha \cdot |\uparrow\rangle_{B2} = f_{BX}(\alpha) \cdot |\psi\rangle_{BX} \qquad (17)$$

Here, too, we defined functions f$_{BX1}$ and f$_{BX2}$, which are measured on branches B$_{C1}$ and B$_{C2}$. Like the previous ones, these are also functions of the possible values of α.

The signal of D$_C$:

$$D_C = D_{C1} + D_{C2} = \frac{1}{2}(\sin^2 \alpha + \cos^2 \alpha) = \frac{1}{2}$$

If the F(XOR) gate performing the interference of the branches B$_{X1}$ and B$_{X2}$ is also a phase-shifting beam splitter, the B$_X$ at the output performing phase-shifting π:

$$B_X: f_X(\alpha) \cdot |\psi\rangle_{BX} = \frac{1}{2}|\uparrow\rangle_{\alpha A}(-\sin \alpha \cdot |\uparrow\rangle_{BX} - \cos \alpha \cdot |\rightarrow\rangle_{BX}) \qquad (18)$$

The detector signal will be:

$$D_X = \frac{1}{4}(-\sin \alpha - \cos \alpha)^2 = \frac{1}{4}(1 + \sin 2\alpha)$$

In the case of high number of photons we can interpret the probabilities as amount of photons and the output of the processing unit Inf will give the ratio of the photon amount detected by detector D$_X$ related to that detected by D$_C$:

$$N_S = \frac{D_X}{D_C} = \frac{1}{2}(1 + \sin 2\alpha) \qquad (19)$$

In both cases treated above N$_S$ is number generated from the amount of the photons registered by the two detectors and is independent on the photon number. It indicates the relative amount of photons interfering in the F(XOR) gate, which on the other hand represents the probabilities with which the photon arrives in B$_{X1}$ and B$_{X2}$ branches, respectively. These probabilities are determined by the orientation of the polarizer used for the measurement on A side relative to the local coordinates on B side.

It is necessary to make an addition to what is discussed here.

From a mathematical point of view, the above derivations were related to the case when the branch A measurement takes place when the member of branch B of the entangled photon pair has already passed through BSHV. From a mathematical point of view, the derivation can also be done in such a way that the A-branch measurement precedes the entry of the B-branch photon into the BSHV. Then writing the state of the entangled photons of arbitrary angle on the orthogonal basis corresponding to the polarizer of branch A (see Appendix A3 and chapter 3.1.1):

$$\frac{1}{\sqrt{2}}(|c_1\rangle|d_1\rangle + |c_2\rangle|d_2\rangle) = \frac{1}{\sqrt{2}}(|\uparrow\rangle_{\alpha A}|\uparrow\rangle_{\alpha B} + |\rightarrow\rangle_{\alpha A}|\rightarrow\rangle_{\alpha B})$$

When the photons pass the polarizer on A side, they have a joint state with the photon in branch B that is passing the BSHV, normalized:

$$\Rightarrow |\uparrow\rangle_{\alpha A}|\uparrow\rangle_{\alpha B} = |\uparrow\rangle_{\alpha A}(\cos \alpha |\uparrow\rangle_{B2} + \sin \alpha |\rightarrow\rangle_{B1}) \qquad (20)$$

When the photons on side A are absorbed, the state of the B side photon becomes, normalized:

$$\Rightarrow |\rightarrow\rangle_{\alpha B} = -\sin \alpha |\uparrow\rangle_{B2} + \cos \alpha |\rightarrow\rangle_{B1} \qquad (21)$$

When comparing Equation (20) with Equation (8) and Equation (21) with Equation (14) it is obvious that we get the same expressions for the photon states independent on the temporal order of the polarizing on side A and beam splitting on side B. This indicates that the two possible sequences are physically the same.

## 4.3 Evaluation of the detection results, intermediate conclusions

As a first step for the analysis, we used a simple phase shifter beam splitter as an F(XOR) gate. The results of the derivations are summarized in Fig. 2 as a function of the possible α values.

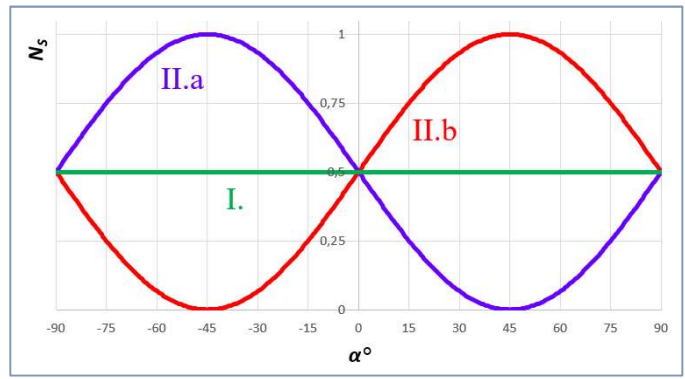

Fig. 2. The effect of the F(XOR) gate implemented with a beam splitter on the $N_S$ relative detection values. **Case I:** No measurement is performed on the branch A; **Case II:** measurement is made with a polarizer with angle α on branch A, and on branch B **Case II.a:** state corresponding to polarization angle α or **Case II.b:** state corresponding to polarization angle α–90° determines the state of the photon to be detected. This diagram represents the statement of the no-communication theorem.

The figure illustrates the effect of the construction in cases I and II. In case I, a constant function emerges in the Inf unit, which is independent of the polarization angle of the photon pair (meaning it remains stable against rotations). In case II, when a measurement is made with an α angle polarizer on branch A, the photon on branch B assumes one of the states corresponding to α or α–90° polarization angle. These two different states result in relative probability values according to curves II.a and II.b in the Inf unit.

The value $N_S$ denotes a statistical value created by a larger quantity of photons corresponding to cases I or II.a or II.b at the detectors and in the unit Inf. In case I (when there is no measurement on branch A), the package of photons produces a stable $N_S$ detection result. However, in case II (when a measurement occurred on branch A), based on current knowledge, we cannot count on the fact that a packet of photons "purely" corresponds to case II.a or II.b. When a measurement is made with an α angle polarizer on branch A, the photon on branch B randomly enters one of the states corresponding to α or α–90° polarization angle, with a 50-50% probability. Therefore, in the measurements of the detectors, cases II.a and II.b appear mixed, and the Inf data processing unit will indicate a value corresponding to the average of these two sub-cases. From Fig. 2, it can be observed that the curves representing cases II.a and II.b are complementary. For any α value, the average values of the functions corresponding to II.a and II.b will be equal to the value of function I (which is constant for any α). This means two things: firstly, in the case of a certain α measurement on branch A, cases I and II cannot be distinguished on branch B, and secondly, for two different α measurements ($α_1$ and $α_2$), these two cases cannot be distinguished. In this interpretation, therefore, the second unit of the setup cannot extract information from the information generated in the first unit. All this corresponds to the assertion of the no communication theorem, a special proof of it.

Conclusion: While it can be proven that information about the entangled photons and their manipulation is generated in the wave function determined by the given construction (in the first unit), extracting this information with detectors in the form of statistical averages (in the second unit) is not obvious. This latter problem was examined using a simple F(XOR) gate implemented with a beam splitter.

The cause of the problem can be intuitively understood. The beam splitter performs a *symmetric* division on both sides. As a consequence, after the symmetric division of two state functions representing complementary probability distributions, their probability averages on both sides of the beam splitter must result in the same values: a 50-50% probability average. If this were not the case, the F(XOR) gate would be able to perform its intended function.

### *4.4 Generalization and an application*

In Appendix A4, we generalize our conclusion made above to the F(XOR) gate implemented with a beam splitter, and prove mathematically that complex systems of elements based on symmetric distribution necessarily lead to the conclusion of the no-communication theorem. Any such F(XOR) gate (for example, operating with combinations of paths created by beam splitters and controlled by phase shifters) produces detection averages that do not depend on the angle α of the measurement performed on branch A.

In Appedix A5, we describe a limited communication application that results from the ability of the construction and the F(XOR) gate implemented with a beam splitter to distinguish between entangled and non-entangled photon packets on branch B. In this way, it is not possible to transmit information between two branches, but meta-information can be obtained on branch B about the existence or absence of the entire system based on entangled photons. That is, about whether an information is shared by entangled photons between two actors or not.

### *4.5 Information processing solutions*

The critical element regarding the possibility of information extraction was the implementation of the F(XOR) gate. With the implementations presented below we demonstrate that information transfer from branch A to branch B is possible, enabling communication using entangled photons without additional information requirement about correlation.

For each type of solution, we examine two specific cases: 1) measurement with a polarizer at a 45° angle on branch A, and 2) measurement with a polarizer at a 90° angle. Both cases fall under the type II. discussed in chapter 4.3.

On branches $B_{X1}$ and $B_{X2}$ before the F(XOR) gate (without beam splitters $BS_1$ and $BS_2$), the previously derived general mathematical formulas (Equations (9) and (15)) apply:

II.a: $f_{BX1}(\alpha) \cdot |\psi\rangle_{BX1} + f_{BX}(\alpha) \cdot |\psi\rangle_{BX2} = |\uparrow\rangle_{\alpha A} \cdot \cos\alpha \cdot |\nearrow\rangle_{BX} + |\uparrow\rangle_{\alpha A} \cdot \sin\alpha \cdot |\nearrow\rangle_{BX1}$

II.b: $f_{BX1}(\alpha) \cdot |\psi\rangle_{BX} + f_{BX}(\alpha) \cdot |\psi\rangle_{BX2} = -\sin\alpha \cdot |\nearrow\rangle_{BX2} + \cos\alpha \cdot |\nearrow\rangle_{BX1}$

(As previously described, the two cases arose from the fact that for any measurement at angle α on branch A, two sub-cases at angles α and α+90° are generated on branch B with a 50-50% probability.)

1) In the case of a measurement at 45° angle on branch A, the two sub-cases are:

II.a(45°): $|\uparrow\rangle_{\alpha A} \cdot \frac{1}{\sqrt{2}} \cdot |\nearrow\rangle_{BX2} + |\uparrow\rangle_{\alpha A} \cdot \frac{1}{\sqrt{2}} \cdot |\nearrow\rangle_{BX1}$

II.b(45°): $-\frac{1}{\sqrt{2}} \cdot |\nearrow\rangle_{BX2} + \frac{1}{\sqrt{2}} \cdot |\nearrow\rangle_{BX1}$

Since the polarizations are aligned, they will not be significant, and thus we will only use notation that contains spatial parameters. We can also get rid of the state of the A photon, since the entanglement broke down during the measurement on branch A. With this simplified notation:

II.a(45°): $f_{BX1}(45°) \cdot |\psi\rangle_{BX1} + f_{BX2}(45°) \cdot |\psi\rangle_{BX2} = \frac{1}{\sqrt{2}} \cdot |B_{X2}\rangle + \frac{1}{\sqrt{2}} \cdot |B_{X1}\rangle$ (22)

II.b(45°): $f_{BX}(45°) \cdot |\psi\rangle_{BX1} + f_{BX2}(45°) \cdot |\psi\rangle_{BX2} = -\frac{1}{\sqrt{2}} \cdot |B_{X2}\rangle + \frac{1}{\sqrt{2}} \cdot |B_{X1}\rangle$ (23)

2) In the case of a measurement at a 90° angle on branch A, the two sub-cases are:

II.a(90°): $|\uparrow\rangle_{\alpha A} \cdot |\nearrow\rangle_{BX}$

II.b(90°): $-|\nearrow\rangle_{BX}$

Transcribed to the simplified notation introduced above:

II.a(90°): $f_{BX}(90°) \cdot |\psi\rangle_{BX1} + f_{BX}(90°) \cdot |\psi\rangle_{BX2} = |B_{X1}\rangle$ (24)

II.b(90°): $f_{BX1}(90°) \cdot |\psi\rangle_{BX1} + f_{BX}(90°) \cdot |\psi\rangle_{BX} = -|B_{X2}\rangle$ (25)

**4.5.1 F(XOR) Gate implementation using a three slit plate for sampling the output of the B branches**

For better clarity, we deviate here from the optical path-based implementation used previously in the construction (although it could certainly be implemented this way as well). We use a three-slit screen to sample the output pattern coming from $B_{X1}$ and $B_{X2}$ branches – this system is behaving as an F(XOR) gate – and then the output arrives at a detector at a specific distance from it. We will not need the control branches $B_{C1}$ and $B_{C2}$ for the calculations, so the beam splitters $BS_1$ and $BS_2$ are not present.

To simplify the proof, we accept idealizations: we assume a large quantity of photons for evaluating detection results; we consider the slits and the branch outputs to have infinitesimally small cross-sections for calculations; we disregard photons that are absorbed by the screen containing the slits.

1) In the case of the 45° angle A branch measurement, the two possible subcases are shown in Fig. 3 (using the notation introduced above).

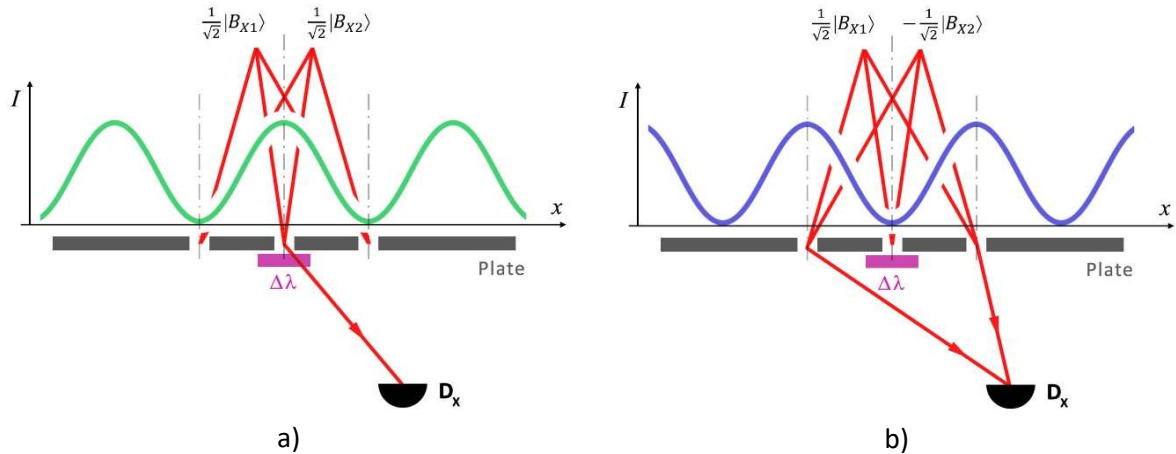

Fig. 3. The effect of an F(XOR) gate implemented by sampling the interference pattern of branches $B_{X1}$ and $B_{X2}$ when measuring with a polarizer at α=45° angle on branch A. In branch B, in **case a)** the state corresponding to the α polarization angle, or in **case b)** the state corresponding to α–90°

polarization angle, determines the state of the photon entering the F(XOR) gate.

Branches $B_{X1}$ and $B_{X2}$ enter the F(XOR) gate containing the slit system. Initially, the $B_{X1}$ and $B_{X2}$ create specific interference patterns on Plate. In case II.a (45°), the photons of the $B_{X1}$ and $B_{X2}$ branches will be in the same phase, while in case II.b (45°), there will be a phase difference of π between them. This results in two distinct interference patterns (green and blue diagrams). The characteristic of the two interference patterns is that they are spatially complementary: where one has maxima, the other will have minima and vice versa. (This complementarity holds for all α angles in general. And if the information processing F(XOR) gate would only be the interference possibility generated by the spatial separation of branches $B_{X1}$ and $B_{X2}$, the no-communication theorem would still hold, as the average of the two complementary functions would yield the same value for all angles.)

The three slits are placed in accordance with the period of interference patterns resulting from the interference of the branch outputs. The central slit is located at the position of the maximum of subcase II.a(45°) and the minimum of subcase II.b(45°). The two outer slits are located at the positions of the two minima in the II.a(45°) subcase and the two maxima in the II.b(45°) subcase. The location of these is determined by the amount of spatial separation of $B_{X1}$ and $B_{X2}$ and their distance to the slits.

2) Let us now consider the case where a measurement at a 90° angle is performed on branch A, in which case the two subcases are shown in Fig. 4 (using the introduced simplified notation).

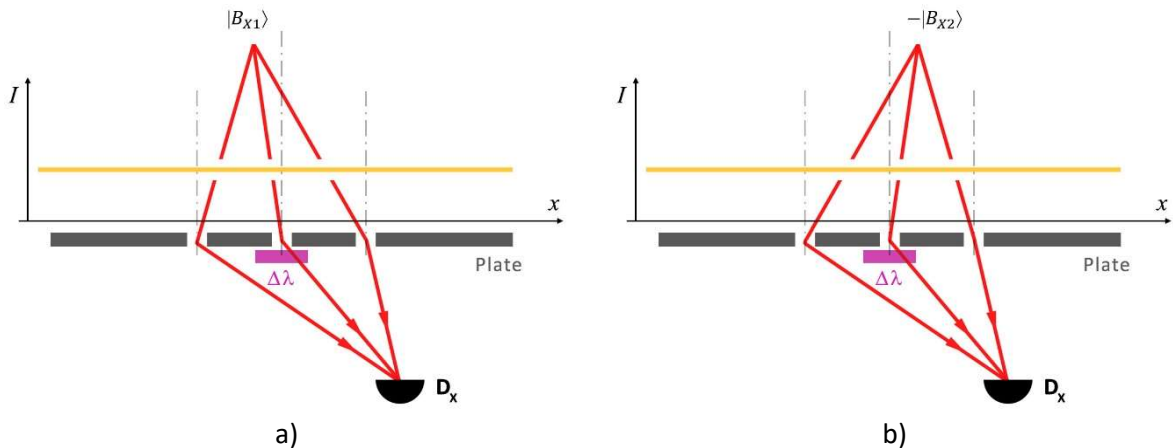

a)                                            b)

Fig. 4. The effect of an F(XOR) gate implemented by sampling the branch output with three slits, when measuring with a polarizer at α=90° angle in branch A. In branch B, in **case a)** the state corresponding to the α polarization angle, or in **case b)** the state corresponding to α–90° polarization angle, determines the state of the photon entering the F(XOR) gate.

These cases are special because one of the superposition components associated with the $B_{X1}$ and $B_{X2}$ branches has a zero value (this can be considered as the photon traveling on one or the other branch). For infinitely small spatial extent of the branches, the signal falling on the slits can be approximated by a constant function (yellow diagrams).

In accordance with the conditions described at the beginning of the chapter, we simulated the optical interferences with a custom program. Appendix A6 details the chosen parameters. In this presentation, the phase shifter Δλ does not play a role. If calculating with a sufficient number of detectors in the plane parallel to Plate, the intensity curves are plotted according to Fig. 5 and Fig. 6. In Fig. 5.a, the curves of subcases II.a(45°) and II.b(45°) for the case 1), while in Fig. 6.a, the curves of subcases II.a(90°) and II.b(90°) for the case 2) are projected onto one diagram. Since the subcases occur randomly in a large number of photon packets, the detectors essentially detect their averages (as explained in Chapter 4.3). Therefore, Fig. 5.b and Fig. 6.b show average curves formed from these subcases.

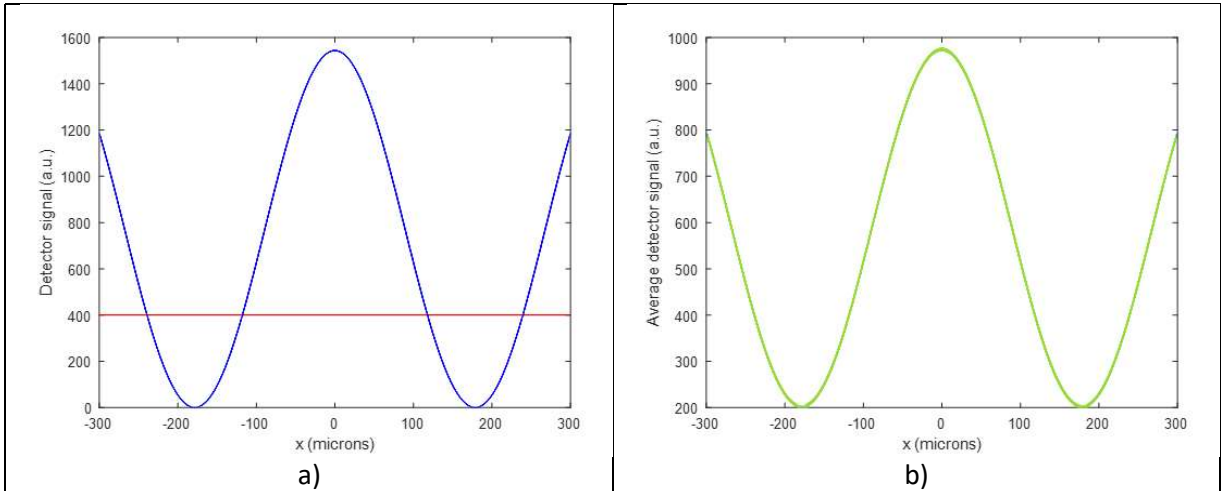

Fig. 5. **a)** Intensity curves of the subcases II.a(45°) (red) and II.b(45°) (blue) of case 1) shown in Fig. 3, in the plane of the $D_X$ detectors; **b)** the average curve (green) of the curves of the two randomly generated subcases gives the effective detections of the detectors.

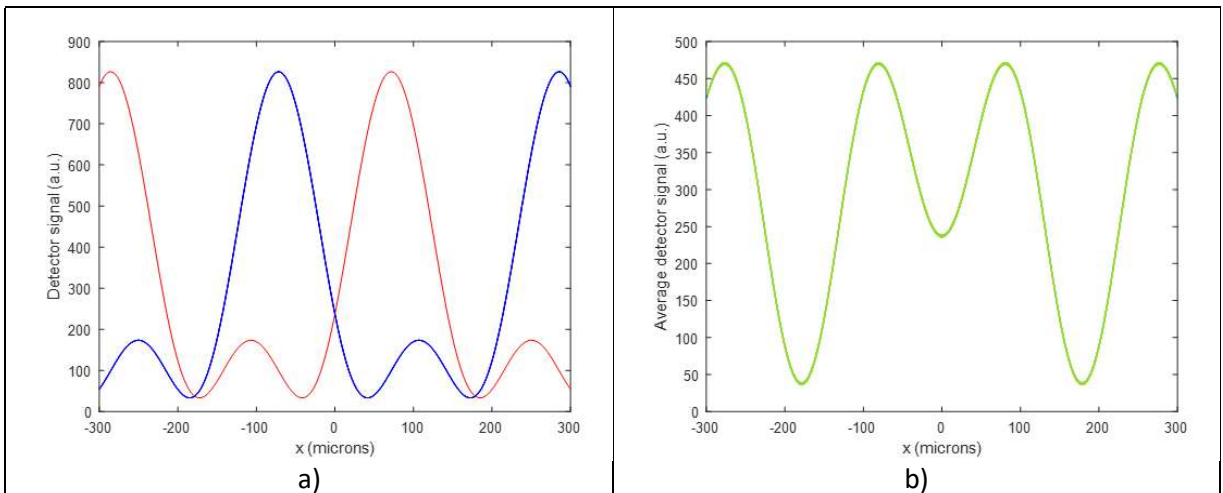

Fig. 6. **a)** Intensity curves of the subcases II.a(90°) (red) and II.b(90°) (blue) of case 1) shown in Fig. 4, in the plane of the $D_X$ detectors; **b)** the average curve (green) of the curves of the two randomly generated subcases gives the effective detections of the detectors.

The curves of Fig. 5.b and Fig. 6.b clearly differ. This means that an F(XOR) gate implementing a plate with three slits to sample the output of the branches leads to different detection results when the measurement on branch A is performed 1) with a polarizer with an angle of α=45°, and when 2) with a polarizer with an angle of α=90°.

Our conclusion: we can consider it proven that there are at least two cases where two different measurement events on branch A lead to distinguishable detection results on branch B.

**A special mathematical proof regarding the degrees of freedom of photons**

In Appendix A7, we provide an easily verifiable direct mathematical proof of the functionality of the F(XOR) gate implemented with the above three slits. At the same time, the proof is also specific, as it shows a more hidden property of the connection created with entangled photons. Using the Δλ phase shifter shown in Figs. 3 and 4 reveals that the measurements performed on branch A can in a certain sense change the degrees of freedom of photons on branch B. Some measuring angles on branch A allow manipulation with the phase shifter on branch B, other measuring angles eliminate this possibility. This may be an effect that has theoretical relevance.

**4.5.2 Two non-trivial asymmetric solutions for F(XOR) gate**

Since the F(XOR) gate is the most critical point of the construction for the information transfer that can be achieved by entangled photons, we also worked on other solutions. Our general conclusion is that there are possible solutions to the problem that we formulated in chapter 4.3 and appendix A4. Different types of solutions must present an asymmetric function. In the II(45°) and II(90°) cases recorded in chapter 4.5, when the photons are in a superposition or non-superposition state in front of the F(XOR) gate, such gates prescribe different types of behavior for the photons (we saw this in the three-slit solution above also). The goal is that the subcases of the two cases lead to different average results in the two cases. We describe two solutions, but since they are not trivial and we also describe possible problems associated with them, we include them in the Appendix.

Our solution described in Appendix A8 is based on the Hong-Ou-Mandel effect. We show that in the II(90°) case, when the photons travel cleanly on one or the other branch, the conditions necessary for the creation of this specific interference can be ensured on a phase-shifting beam splitter. Photons in the II(45°) superposition state do not produce detection results typical of Hong-Ou-Mendel interference.

In Appendix A9, we outline the function of a hypothetical interference space (IS), in which the photons from one or the other branch of the II(90°) case continue their journey unhindered and without interference, however in the two subcases of the II(45°) case, the superposition state photons form constructive or destructive interference. The detection average in the second case will differ from the average formed in the first case. This F(XOR) gate deserves attention because, just like in the special proof presented in Appendix 7 belonging to chapter 4.5.1, the degree of freedom of the photons changes in the two cases as a result of the A branch measurement.

These examples also show that there is no reason to rule out the possibility of further, even more effective solution types.

# 5. Summary. The faster than light communication

We presented a model from the perspective of quantum information with three possible solutions. Based on quantum mechanical analysis, we consider the possibility of communication faster than the speed of light to be proven. The entangled pair of photons is sent to two distant participants, and we have shown that different measurements performed on one member of the pair may be distinguished by appropriately measuring the other member of the pair. This distinction leads to the possibility to retrieve information encoded into the chosen measurement.

We provided a broader framework for measurement conditions than used in the proof of the no-communication theorem. This framework was based on the introduction of weak measurement. On one branch of the entangled photons (B), we suggest a construction that measure the quantum state of the entangled photons using polarization dependent spatial separation. A polarization dependent beam splitter connects two types of superposition states: those belonging to polarizations and spatial positions. The possible values of these two superposition states (probability amplitudes) are correlated. The polarization state of the photon in the receiver side (B) is determined by the polarization of its entangled pair at the sender (A), whereas this polarization state is transformed into spatial state. The quantum information carried in the spatial superposition state of individual photons can be extracted exhausting spatial interference. Interference occurs only in a certain polarization state, a special F(XOR) gate and detectors make distinction between interfering and non-interfering photons. The signal processing unit forms a relative quantity of the ratios of non-interfered and interfered photons. The task was to create an F(XOR) gate that assigns different detection quantities on the receiver (B) branch to different measurement events on the sender (A) branch. The solution is not trivial, as shown in the discussion of the problem of complementarity.

We came to the conclusion that, in general, F(XOR) gates with a so-called symmetric function cannot solve this problem. The proposed F(XOR) gate implementations create interference in a certain way asymmetrically with the superposition components of different states. We presented three different possible solutions. One type of gate introduced asymmetric sampling with three additional slits, the other type of gate asymmetrically distinguishes superposition and non-superposition state photons using the Hong-Ou-Mendel effect, and a third theoretically determined gate exhausts the asymmetric outcome when the spatially separated superposition state is recombined under zero degrees. We do not rule out that such asymmetrically functioning implementations could be realized in other ways that offer greater efficiency.

Overall: we outlined a theoretical construction enabling quantum communication, which bases the information transfer not on the correlation of definite value measurement values of quantum states (which was disproved by the no-communication theorem), but on the correlation of indefinite value superposition states produced on the basis of weak measurement. Strictly speaking, the members of one branch of the photon pairs (B) do not perceive the change of the photons of the other branch (A) (through correlation); instead, they perceive changes in the system formed by the photon pairs, of which they are also parts. From a communication theory perspective, we interpret this not as the transaction of information (as in information theories) but as participation in the information system (the participation theory (PTC) was outlined in Appendix A5).

The outlined quantum communication solution does not imply immediate information transfer. The condition for this would be that individual photons carry information. However, this is not possible; information can only be provided by the statistical average of a larger number of detected photons. If the "time window" for the passage of the appropriate number of photons through the construction is less than the time it takes for light to travel the distance between the photon pairs, then faster-than-light communication is achieved.

**General technical notes:**

1. The previous description of information sharing by entangled photons was based on the ideal case when the photon source So emits only such photon pairs. In practice, such a pure photon source is not available, entangled photon pairs are generated in a rather small proportion compared to the amount of non-entangled photons. This means that in the physical implementation of the construction, the different signals will be less contrasting in the detector system. This presents a practical obstacle to reliable application. However, the entangle photon pair generation possibilities have advanced significantly, and further progress is expected. Thus, in the event of the realization of the construction, achieving higher efficiency can also be predicted.

2. The efficiency of information sharing achievable by the construction (the relative contrast of the signals) may also depend on how the choice of polarization bases between branches A and B relates to each other. One option is obviously for the operators of the two branches to agree these via a classical channel beforehand. Another solution, however, is to calibrate the entire system without such an auxiliary channel. The basis for this is the observation that in the case where no measurement occurs on branch A (case I.), constant detection values are created in the system of branch B (see chapters 3.1.1 and 4.1): the system is immune to rotations. (This could even be one value of a binary code.) Relative to this, the base of the B branch system can be calibrated so that the largest signal deviation occurs when an angle measurement is made on branch A (case II.).

In the implementation of the F(XOR) gate using three-slits (chapter 4.5.1), we also described a specific calibration option using a phase shifter.

3. Since solutions based on communication via entangled states do not require the insertion of a classical auxiliary channel, a question may arise regarding the synchronization of signals produced by measurements of specified durations on branch A and detections of photon packets of specified durations on branch B. Without coordination, the boundaries of the signal durations on branch A will not coincide with the processing time windows on branch B. As a result, two consecutive signals can mix within a time window. One solution could be to make the duration of a signal on branch A multiple times longer than the processing time window on branch B. Thus, when a signal changes on path A, a mixed signal is placed in one or two time windows on branch B, which is followed by several time windows with clear signals. The signal boundaries will be somewhat "blurred," but the uniform values of consecutive time windows will indicate the signal value.

4. As a question of principle, it can be raised that if a message corresponding to some kind of coding is successfully decoded on branch B, the actor in branch B ultimately cannot gain complete certainty whether this message originated from branch A using the entangled photon system. Ultimately, this can only be achieved through the application of a classical auxiliary channel, where coincidence testing and comparison of the detection values generated on the two branches take place. This is the logic of current informational applications of entangled photons. We would describe the problem with a non-quantum communication analogy. Assume that Bob is waiting for a message from Alice over a simple Morse-coded radio channel. Based solely on the received message, he can never be sure whether this message is the result of radio signals generated by a solar flare, which, in a highly unlikely manner (but not with zero probability), form a meaningful sequence, or whether they are from a third party trying to deceive Bob, or have been created in some other way. However, all these possibilities do not negate the fact that communication between Alice and Bob is possible.

# APPENDICES

## A1. Entangled particle pair

Quantum entanglement is one of the most fundamental phenomena in quantum mechanics that has no counterpart in classical physics, and cannot be interpreted using classical physical concepts and mathematical formalisms.

The phenomenon of quantum entanglement generally occurs when particles within a group interact with each other in a specific way, come into spatial proximity in a defined manner, or are created in a specific way. While the particles, during the development of their quantum states, move to macroscopically interpretable distances from each other in space, (more precisely, become detectable at macroscopically interpretable distances), the quantum state of individual particles cannot be described independently of the states of the other particles. [12]

Entanglement implies a relationship between particles such that in the entangled quantum state, these particles form a system and cannot be described otherwise. The quantum state that creates entanglement is fundamentally based on indeterminacy: the states of the particles comprising the system do not possess definite values. The third characteristic is expressed in the measurement event and its outcomes: in phenomena indicating entanglement, measurement results contain correlations that differ from random statistical averages.

The wave function describing the entanglement corresponds to these characteristics. A four-dimensional wave function describable in the so-called configuration space, involving two two-dimensional systems [13]:

$$\psi = |v\rangle|w\rangle = r|a_1\rangle|b_1\rangle + s|a_1\rangle|b_2\rangle + t|a_2\rangle|b_1\rangle + u|a_2\rangle|b_2\rangle \qquad (a1)$$

The second part of the formula expresses the mathematical interconnection along the states $|v\rangle$ and $|w\rangle$ of the two particles, while the third part describes the indeterminacy of one particle in states $|a_1\rangle$ and $|a_2\rangle$, and the indeterminacy of the other particle in states $|b_1\rangle$ and $|b_2\rangle$.

The sum of the statistical probabilities:

$$r^2 + s^2 + t^2 + u^2 = 1 \qquad (a2)$$

This formula expresses that the sum of the probabilities of measurement events associated with each quantum state must equal 1.

The relationship between the probability amplitudes is:

$$ru \neq st \qquad (a3)$$

This formula establishes the criterion for phenomena carrying entanglement. From this condition, it follows that measurement results contain correlations that deviate from random statistical averages.

Pure Bell states can be considered as special entangled states, where:

$$ru = 0 \text{ or } st = 0$$

Their speciality lies in the fact that as a result of this condition, the correlations between the measurement results of the two particles are deterministic, meaning 100% probability. Their peculiarity arises from the existence of a "global" characteristic of the entire system among the possible combinations of states expressing the particles' indeterminacy, which remains constant:

Ha $ru = 0$ then $st = ct$

Ha $st = 0$ then $ru = ct$         (a4)

In our discussions and potential implementations, entangled photons in such pure Bell states will be involved. Our goal is to gain information about this "global" characteristic or its manipulation at a "local" level, i.e., concerning only one of the particles. As we will elaborate, this could serve as the core for information transmission and communication achievable through entangled particles.

## A2. Quantum nonlocality and quantum entanglement

The concept of non-locality has been present in some form since the birth of quantum mechanics. Essentially, it is closely related to the most fundamental feature of quantum mechanics, the collapse of the wavefunction. In the Copenhagen interpretation, quantum entities exhibit wave-particle duality - upon measurement, the wavefunction describing quantum states collapses, and they are detected as particles. The pre-measurement wave characteristics spread out in space, but the collapse is instantaneous, and it seems as if some sort of coordination spreads faster than light between distant points in space. However, quantum mechanics does not describe this process. In the early days of quantum mechanics, this was considered more of an interpretational difficulty.

Very similarly, but in principle the same, the feature of non-locality is demonstrated by the so-called EPR paradox, which describes two entangled particles that can be spatially separated but belong to the same wavefunction. When a measurement is made on one particle, the entire wavefunction collapses in a specific way, and the other particle ends up in a corresponding state. This results in correlated measurement outcomes for the two particles (completely in the case of pure Bell states). The authors describing this problem originally interpreted this type of correlation as a kind of action at a distance, which they saw as paradoxical because it contradicts the laws of propagation of effects across localities, which exclude distant action at a distance. [14] However, the correlation simply emerged from quantum mechanical derivations, leading them to believe that the theoretical system was incomplete. Therefore, they postulated hidden parameters that would uphold the principle of locality. This implies that the correlation does not occur when the two particles are far apart and measured, but is predetermined by the hidden parameters. This idea sharply contrasts with interpretations that argue for the completeness of quantum physics. The contradiction was highlighted by a theorem known as the Bell inequality [15], showing that theories postulating predetermined correlations of measurement results are incompatible with those that do not postulate them as predetermined. Based on the theorem, experimental implementations could be built to answer the question, and as we know, they did not support theories based on locality (one of the strongest of many experimental justifications: [16]).

It is therefore considered well-founded to conclude that so-called locally realistic theories are incorrect if realism means that relevant physical properties are independent of measurement or observation (i.e., they are predetermined before these events), and locality means that the measurement separately of two entangled entities does not influence each other (as they are predetermined before measurement). [17] (We are not specifically concerned with hidden variable theories at this time.)

This poses a problem because the conditions of local realism would be compatible with special relativity theory and the speed-of-light constraint on the propagation of effects set here. However, rejecting these theories means that due to the pre-measurement quantum mechanical indeterminacy and uncertainty, the correlations between measurements are not deterministically predetermined.

Thus, this correlation must occur in the events of measurements, which necessarily implies that separate measurements determine each other. If the measurements occur simultaneously, this determination is immediate – the speed-of-light limit of special relativity theory cannot be ensured. These theories are theories that assume non-locality. [18]

The principle of non-locality remains a topic of widespread debate today. Therefore, it proves to be a fundamental question whether the correlation actually arises in events ensured by measurement settings independent of the quantum states. Such questions include whether, if measurements on entangled entities determine each other, these can be considered genuine interactions. In this case, can the logical requirement of causality be ensured? [19]

## A3. The BSHV for entangled photons

An entangled pair of photons emerges from the photon source, which is characterized by a pure Bell state composed of $(|a_1\rangle, |b_1\rangle)$ and $(|a_2\rangle, |b_2\rangle)$ states written in an orthonormal basis. It is well known that rotating the normalized base by an angle α (arbitrary) does not influence the form of the entangled state written using the polarization state vectors [4], for our basis determined by the orientation of $P_H$ and $P_V$:

$$\frac{1}{\sqrt{2}}(|a_1\rangle|b_1\rangle + |a_2\rangle|b_2\rangle) = \frac{1}{\sqrt{2}}\left(|\uparrow\rangle_A|\uparrow\rangle_B + |\rightarrow\rangle_A|\rightarrow\rangle_B\right) \qquad (a5)$$

The invariability of the state is expressed in the invariability of the coefficients of the state vectors, representing their probability amplitudes. When the relation between the coefficients also expresses the relative phase between the possible states, we can emphasize that this phase remains constant when we rotate the basis (the polarization). We can handle this as an information characteristic to the whole system.

When we take into account the spatial division happening in BSHV, the compound state will be the following within the bases determined by the splitter:

$$\frac{1}{\sqrt{2}}(|a_1\rangle|b_1\rangle + |a_2\rangle|b_2\rangle) =$$

$$= \frac{1}{\sqrt{2}}\left(\cos\alpha|\uparrow\rangle_A + \sin\alpha|\rightarrow\rangle_A\right)\left(\cos\alpha|\uparrow\rangle_{B2} + \sin\alpha|\rightarrow\rangle_{B1}\right)$$

$$+ \frac{1}{\sqrt{2}}\left(-\sin\alpha|\uparrow\rangle_A + \cos\alpha|\rightarrow\rangle_A\right)\left(-\sin\alpha|\uparrow\rangle_{B2} + \cos\alpha|\rightarrow\rangle_{B1}\right) =$$

$$= \frac{1}{\sqrt{2}}\left(|\uparrow\rangle_A|\uparrow\rangle_{B2} + |\rightarrow\rangle_A|\rightarrow\rangle_{B1}\right)$$

## A4. The general problem of information processing with symmetric function F(XOR) gates

With Fig. 2, we have shown that if we try to obtain information from the information created in the first unit of the construction in the second unit with an F(XOR) gate implemented with a simple beam splitter, this cannot succeed. Below we give the general conditions for all solutions that do not allow information processing – which thus correspond to the conclusion of the no communication theorem.

The general case can be written for all modifiers forming the F(XOR) gate or their complex system, where their effect can be expressed in the form of linear relationships resulting in *symmetric* divisions (such modifiers are e.g. beamsplitters, combined with scalable phase shifters, which affect the ratios of the superposition components of branches $B_{X1}$ and $B_{X2}$):

$$B_X: f(\ldots x_i, \ldots) \cdot b_1 \cdot |\nearrow\rangle_{BX} + f(\ldots x_m, \ldots) \cdot b_2 \cdot |\nearrow\rangle_{BX} \qquad (a6)$$

where

$b_1$ and $b_2$ – probability amplitudes for branches $B_{X1}$ and $B_{X2}$ before the F(XOR) gate;

*f(… $x_i$ …)* and *f(… $x_m$ …)* – they express the system of modifiers affecting the original probability amplitudes $b_1$ and $b_2$ placed on the $B_X$ branch. Mathematically, they are created by subtracting $b_1$ and $b_2$ from the wave function that can be written on the $B_X$ branch, and these these functions come out as multipliers.

We substitute the characteristic probability amplitudes corresponding to the three cases (I., II.a, II.b) in place of $b_1$ and $b_2$ in the written general formula:

case I.: $B_X: \dfrac{1}{4}\big(f(\ldots x_m, \ldots) \cdot |\nwarrow\rangle_A |\nearrow\rangle_{BX} + f(\ldots x_m, \ldots) \cdot |\nearrow\rangle_A |\nearrow\rangle_{BX}$
$\qquad\qquad + f(\ldots x_i, \ldots) \cdot |\nwarrow\rangle_A |\nearrow\rangle_{BX} - f(\ldots x_i, \ldots) \cdot |\nearrow\rangle_A |\nearrow\rangle_{BX}\big)$

$= \dfrac{1}{4}[f(\ldots x_m, \ldots) + f(\ldots x_i, \ldots)] \cdot |\nwarrow\rangle_A |\nearrow\rangle_{BX} + \dfrac{1}{4}[f(\ldots x_m, \ldots) - f(\ldots x_i, \ldots)] \cdot |\nearrow\rangle_A |\nearrow\rangle_{BX}$

$N_S = \dfrac{D_X}{D_C} = \dfrac{1}{8}[f(\ldots x_m, \ldots) + f(\ldots x_i, \ldots)]^2 + \dfrac{1}{8}[f(\ldots x_m, \ldots) - f(\ldots x_i, \ldots)]^2$

$\qquad\qquad = \dfrac{1}{4}[(f(\ldots x_m, \ldots))^2 + (f(\ldots x_i, \ldots))^2] \neq f(\alpha)$

case II.a: $B_X: \dfrac{1}{2}|\uparrow\rangle_{\alpha A}\big(f(\ldots x_m, \ldots) \cdot \cos\alpha \cdot |\nearrow\rangle_{BX} - f(\ldots x_i, \ldots) \cdot \sin\alpha \cdot |\nearrow\rangle_{BX}\big)$

$N_S = \dfrac{D_X}{D_C} = \dfrac{1}{2}[f(\ldots x_m, \ldots) \cdot \cos\alpha - f(\ldots x_i, \ldots) \cdot \sin\alpha]^2$

case II.b: $B_X: \dfrac{1}{2}\big(-f(\ldots x_m, \ldots) \cdot \sin\alpha \cdot |\nearrow\rangle_{BX} - f(\ldots x_i, \ldots) \cdot \cos\alpha \cdot |\nearrow\rangle_{BX}\big)$

$N_S = \dfrac{D_X}{D_C} = \dfrac{1}{2}[-f(\ldots x_m, \ldots) \cdot \sin\alpha - f(\ldots x_i, \ldots) \cdot \cos\alpha]^2$

Average of cases II.a and II.b:

$\overline{N_S} = \dfrac{1}{4}[(f(\ldots x_m, \ldots))^2 \cdot (\cos^2\alpha + \sin^2\alpha) + (f(\ldots x_i, \ldots))^2 \cdot (\sin^2\alpha + \cos^2\alpha)] \neq f(\alpha)$

$N_S \text{ (case I.)} = \overline{N_S} \text{ (case II.a, case II.b)} \qquad (a7)$

Overall, we came to the conclusion that the symmetric effect modifiers forming the F(XOR) gate have an effect on the outcome of interferences created on the basis of spatial relations. However, these effects ultimately do not depend on the relative phase characterizing the entanglement phenomena (which is expressed by the wave function of case I), nor on the parameters of the measurement performed on them (the measurement performed on branch A on the setting of the α value polarizer in case II). At the same time, the results of cases I and II corresponding to the effects of the modifiers lead to equal values. All this means that the F(XOR) function based on this type of modifiers does not allow the extraction of system information. This also means that the F(XOR) gates corresponding to the prescribed general formula do not allow communication between branches A and B.

## A5. Discrimination between entangled and unentangled photons – a limited communication application

The field of communication sciences explores forms and conditions of communication that are not solely based on direct information transfer. In these theories and models, information is not only encoded in the actual content of the transaction but can also exist or emerge within the communication context itself. While these issues are traditionally addressed by the humanities, here we present the physical conditions for such an interpretation.

Let's consider the following communication scenario. Source (So) can send information to both branch A and branch B. The actual information is not encoded in the polarization direction of the photons, but, for example, in the temporal extent of light pulses (such as Morse code). Some information can only be received by the participant on branch B (Bob), while other information can be accessed by both branches (Alice and Bob). Bob needs this meta-information. Let's assume that this can be categorized into two values: 0 – if the sent information is not accessible on branch A (Alice cannot possess it), 1 – if the sent information is accessible on branch A (Alice can possess it). These two values thus indicate the type of information for the participant on branch B, depending on whether an information is jointly owned or not.

The implementation could be as follows. If the information is not shared, the source So radiates coherent light pulses with α polarization only towards Bob. If the information is shared, the source So radiates pulses formed by entangled photons according to the polarization towards both Alice and Bob (the basis of entangled photons can be determined by an α angle). Based on Fig. 2, at the detection on branch B (Bob), the package of non-entangled or single photons corresponds to some α value of curve II.a, while the entangled photons will result in the case of curve I. According to the $N_S$ value, in the first case, 0 can be produced, and in the second case, 1 can be produced, categorizing the type of information. (We do not want to assume further details, but we can still state that in this situation it is not a condition that So has information about the type of information that can be used to label the basic information.)

Such situations, with their physical conditions as described, can be interpreted as communication, for example, by the Participation Theory of Communication (PTC). [20] According to this theory, communication is not only when the sender directly transmits information to the receiver but also when information is created at one agent about information held or missing by another (or even multiple) agents. The theory interprets this as the agent actually gaining information about the entire situation, of which other agents are determinants. The agent can be part of a more general context. Similar questions are discussed in certain systems theories, where the functions of elements must be defined as parts and representatives of the relationships and wholeness of a system.

## A6. Gate implementation using a three slit plate for sampling the output of the B branches

Here we present simulations performed in a custom program modeling optical interference in a classical approach for the cases shown in Figs. 3 and 4. In the plane of the $D_X$ detector shown in the figures, which is parallel to Plate, a line of detectors is used, which provide the values of the light intensities at each point. Based on the markings in Fig. A7.1 (see Appendix A7), we used the following parameters:

λ = 500 nm; a = 2 mm; $d_1$ = 1 m; $d_2$ = 10 cm

**Case 1) Measurements are performed with a polarizer with an angle of α=45°on branch A (**Fig. 3)

The intensity distribution in front of the screen containing the three slits (Plate) depends on the phase difference between the outputs of the branches $B_{X1}$ and $B_{X2}$. The distance of the slits on the three slit-screen is 70 microns, but it is matched to the period of the interference pattern generated by the two output. The slits are placed at a distance of a half period of this pattern, as shown in Fig. A6.1.

Fig. A6.1a shows the intensity distribution when the phase difference is 0 (subcase II.a(45°)), and Fig. A6.1b shows the intensity distribution with a phase difference of π (subcase II.b(45°)).

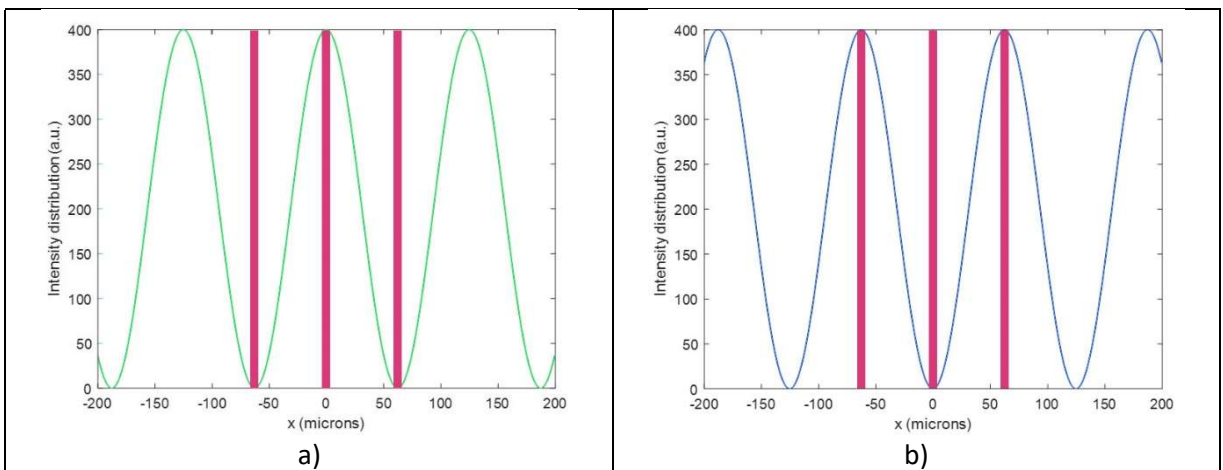

Fig. A6.1. Intensity distribution in front of the screen containing three slits (Plate) produced by two outputs of the branches from photons with 0 phase difference (a) and photons of π phase difference (b). The purple bars represent the locations of the slits on the screen.

The distribution in the plane of the detector after the screen with three slits (Plate) is calculated either from the beam passing through the central slit (Fig. A6.1a) or the interference of the light beams passing through the two side slits (Fig. A6.1b). The calculated intensity diagrams are shown in Fig. 5.

The intensity distribution in the plane of the detector determines the detector signal. When the phase difference between branches $B_{X1}$ and $B_{X2}$ is 0 (subcase II.a(45°)), than their interference at the three slit plane (Plate) results in a peak in the central slit and zeros in the two side slits. Hence the intensity distribution in the detector plane is a constant (green in Fig. 5a), approaching the diffraction pattern of the infinitesimally thin central slit. When the phase difference between branches $B_{X1}$ and $B_{X2}$ is π (subcase II.b(45°)), their interference pattern results in two maxima at the side slits, and zero in the central slit (Plate). Hence the intensity distribution in the detector plane is due to the interference of the beams passing the two side slits (blue in Fig. 5a).

Since the two cases displayed by the two curves in Fig. 5a. (subcase II.a(45°) and subcase II.b(45°)) are mixed due to the reasons described in the main text, the effective detection of the detectors in the case of a large number of photons is the average of these two curves represents. The calculated average curve is shown in Fig. 5b.

Since the two cases displayed by the two curves in Fig. A6.5 (subcase II.a(90°) and subcase II.b(90°)) are mixed due to the reasons described in the main text, the effective detection of the detectors in the case of a large number of photons is the average of these two curves represents. The calculated average curve is shown in Fig. 6b.

Fig. 5b. showing the effect of case 1) and Fig. 6b. showing the effect of case 2) are definitely different. This means that the branch B detector system can distinguish between the two types of branch A measurements resulting in the two cases.

## A7. A special mathematical proof regarding the degrees of freedom of photons

We place a phase shifter after the middle slit of Plate, as shown in Fig. 3. and Fig. 4. We examine how this element can affect the detection values of the cases 1) and 2) defined in chapter 4.5.

In subcase II.a(45°) of case 1) photons only pass through the central slit (Fig. 3.a), creating a single path, making the phase of the photons irrelevant to the measurement result of the $D_X$ detector (the phase shifter does not affect the result). In subcase II.b(45°) of case 1) photons do not pass through the central slit (Fig. 3.b), which also does not affect the detection results. The average of the two sub-cases obviously does not depend on the presence of the phase shifter. In both sub-cases (II.a(90°) and II.a(90°)) of case 2) for, the superposition components of the photons in all three slits are non-zero for most settings, so the phase shifter placed at the central slit will have an effect on the interference formed at the $D_X$ detector in both sub-cases, influencing the detection results.

Below, we show in detail the calculations. First, based on the subcases II.a(45°) and II.b(45°) of case 1 (Fig. 3), we specify the spatial position of the three slits on Plate. And then we calculate the interference obtained on the screen after the three-slit and its local detection value at the $D_X$ detector for subcases II.a(90°) and II.b(90°) of case 2) (Fig. 4). In the latter case, we prove that the phase shifter has an effect on the detection values.

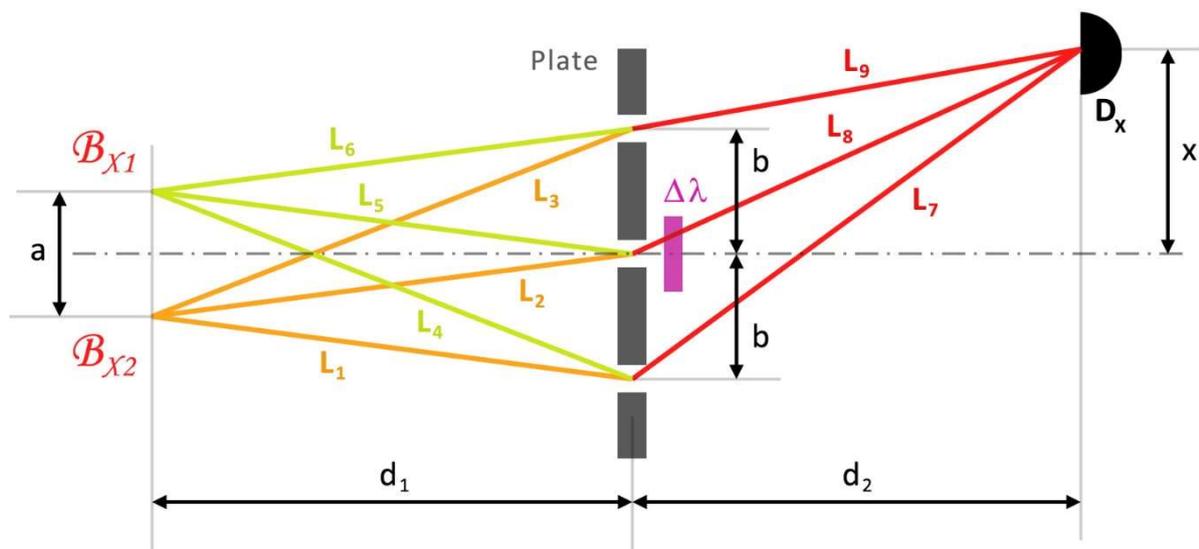

Fig. A7.1. Parameters of the construction of an F(XOR) gate implemented by combining two + three-slit

According to the description referring to Fig. 3, the spatial positions of the three slits on Plate must be calculated based on the interference pattern of the light output from the branches.

Due to the sensitivity of the question outlined in the article, we do not use the rounding formula for Fraunhofer diffraction, but give an exact formula (for the case when the size of the slits approaches zero).

If the waves in the two output of branches are in phase (this is subcase II.a(45°)), the main maximum location of the interference will be on the midline between the two outputs. To calculate the location of the interference minima, consider the following condition:

$$L_3 = L_6 + M\lambda$$

where

$$M = m + 0.5$$

and

$$m = 0; 1; 2 \ldots$$

specifies the sequence number of the minima in the interference pattern.

$L_3$ and $L_6$ paths expressed as a function of the setting parameters:

$$\sqrt{d_1^2 + \left(b + \frac{a}{2}\right)^2} = \sqrt{d_1^2 + \left(b - \frac{a}{2}\right)^2} + M\lambda$$

By derivation, the location of the interference minima:

$$b = \sqrt{\frac{4M^2\lambda^2 d_1^2 - M^4\lambda^4 + M^2\lambda^2 a^2}{4a^2 - 4M^2\lambda^2}}$$

Based on this, the length of all the paths between the locations of the two output of branches and the two interference minima and the main maximum on Plate can be specified.

$$L_1 = \sqrt{d_1^2 + \left(b - \frac{a}{2}\right)^2}; \quad L_2 = \sqrt{d_1^2 + \left(\frac{a}{2}\right)^2}; \quad L_3 = \sqrt{d_1^2 + \left(b + \frac{a}{2}\right)^2}$$

$$L_4 = \sqrt{d_1^2 + \left(b + \frac{a}{2}\right)^2}; \quad L_5 = \sqrt{d_1^2 + \left(\frac{a}{2}\right)^2}; \quad L_6 = \sqrt{d_1^2 + \left(b - \frac{a}{2}\right)^2}$$

If the position of the three slits on Plate coincides with the maximum and two minima of the projected interference pattern, i.e. the "b" parameter is known, the length of the paths between the positions of the three slits and the location of the $D_x$ detector:

$$L_7 = \sqrt{d_2^2 + (x + b)^2}; \quad L_8 = \sqrt{d_2^2 + x^2}; \quad L_9 = \sqrt{d_2^2 + (x - b)^2}$$

Now, the second step will be to calculate the interferences and detection values for case 2) (Fig. 4), which has two subcases: II.a(90°) and II.b(90°) (in the derivation we will use distinguishing indices "a" and "b").

Based on the total lengths of the paths of the light rays from outputs of the branches to the $D_X$ detector, the phase values (in degrees) upon arrival at the detector can be calculated:

$$\varphi_{a1} = \frac{360}{\lambda}[(L_1 + L_7)]; \quad \varphi_{a2} = \frac{360}{\lambda}[(L_2 + L_8 - \Delta\lambda)]; \quad \varphi_{a3} = \frac{360}{\lambda}[(L_3 + L_9)];$$

$$\varphi_{b1} = \frac{360}{\lambda}[(L_4 + L_7)]; \quad \varphi_{b2} = \frac{360}{\lambda}[(L_5 + L_8 - \Delta\lambda)]; \quad \varphi_{b3} = \frac{360}{\lambda}[(L_6 + L_9)]$$

The amplitudes resulting from the interference in the two subcases:

$$A_a = \sin\varphi_{a1} + \sin\varphi_{a2} + \sin\varphi_{a3}$$

$$A_b = \sin\varphi_{b1} + \sin\varphi_{b2} + \sin\varphi_{b3}$$

As described in chapter 4.3, in a package containing a large number of photons, subcases II.a(90°) and II.b(90°) will be present with equal probability, mixed, therefore an average must be calculated for the characteristic of the detection result:

$$\overline{A^2} = \frac{1}{2}(A_a^2 + A_b^2)$$

The relative effect of the different settings of the phase shifter ($\Delta\lambda_1$ and $\Delta\lambda_2$):

$$Phase\ shifter\ effect = \frac{\overline{A^2_{\Delta\lambda1}}}{\overline{A^2_{\Delta\lambda2}}} \neq 1 \qquad (a8)$$

The relation given in this way represents the change of light intensities as a function of the phase shifter settings (which cannot be 1, as this would mean that the phase shifter has no effect).

In the case of $\Delta\lambda2 = 0$ and $\Delta\lambda1 \neq 0$, the relationship shows the effect of a given phase shifter compared to the case without a phase shifter.

As an example, with arbitrarily chosen parameters and settings:

$\lambda$ = 0.7 μm; a = 777 μm; m = 0; $d_1$ = 3333333 μm; $d_2$ = 777777 μm; x = 333333 μm;

and $\Delta\lambda_1$ = 0,3 μm; $\Delta\lambda_2$ = 0

The result that comes out:

$$Phase\ shifter\ effect = 2,466990518$$

The calculations show that the effect of the phase shifter is preserved in the average of the two subcases of case 2). Identifying the function of the F(XOR) gate with the *Phase shifter effect indicator* defined there:

$$f_X(45°) \cdot |\psi\rangle_{BX} = Phase\ shifter\ effect = 1$$

$$f_X(90°) \cdot |\psi\rangle_{BX} = Phase\ shifter\ effect \neq 1$$

So, we have found two cases that fundamentally differ in terms of whether the phase shifter integrated into the slit system of the F(XOR) gate influences the detection results or not. In case 1), it does not, while in case 2), it does. This alone is sufficient to assert that on branch B, cases 1) and 2) can be distinguished, which resulted from two different measurements on branch A. In practical application, the role of the phase shifter may be that while the results from case 1) are given, in case 2) the effect of the phase shifter can be set so that the detection values differ as much as possible from the former.

In practical applications, the role of the phase shifter can be that, while the detection values resulting from case 1) remain constant in contrast to the variable values of the phase shifter, in case 2) the effect of the phase shifter can be adjusted to maximize the difference in detection values from the former. The significance of this calibration is that the bases of the polarization measurements can be coordinated even without the use of a classical channel for communication between branches A and B.

There is a peculiar consequence of the fact that on branch B, changing the phase shifter has no effect on the detection results in case 1), while it does in case 2). The manipulations on branch A – forming either case 1) or case 2) – are thus able to change the number of degrees of freedom of photons in branch B in a certain sense. This effect appears to be more than what follows from the correlation of the entangled photon states – even though it required a specific design of the F(XOR) gate. The potential theoretical significance of this cannot be fully assessed at this point.

Our conclusion: we can consider it proven that there are at least two cases where, as a result of two different measurement events on branch A, different detection results can be distinguished on branch B.

## A8. F(XOR) gate using the Hong-Ou-Mendel effect

Our solution based on the Hong-Ou-Mandel effect is technically more challenging to implement, but in principle, it is capable of extracting the information generated in the first structural unit of the construction. The Hong-Ou-Mandel effect is a two-photon interference effect, where photons with the same properties enter through two inputs of a beam splitter – in the beam splitter, they are indistinguishable – and after a specific destructive interference, only the same output options remain for the two photons. Both photons exit one *or* the other side of the beam splitter and have zero probability of exiting on one *and* the other side. [21]

The F(XOR) gate is therefore implemented with a beam splitter, one half of which performs a π phase shift. So far, this aligns with what has been summarized in chapter 4.3. For the Hong-Ou-Mendel effect to occur, one of the basic conditions for the indistinguishability of photons is that they enter the beam splitter at the same time. In the construction, this is ensured with two elements: a photon source that emits photons with a time interval difference of Δt (So(Δt)), and a delayer that delays the arrival of photons traveling in the $B_2$ branch to the beam splitter by a time interval Δt (Del) (Fig. A8.1). (The representation of the photon source with an icon resembling a camera aperture symbolizes that photons arrive in adjustable doses, technically, e.g. can be implemented with a pulsed laser.) The combined effect of these two elements is as follows: a photon that exits the photon source at any given time and travels in the $B_2$ branch, and another photon that exits the photon source Δt time interval after the previous photon and travels in the $B_1$ branch, will simultaneously arrive at the two surfaces of the $BS_X$ beam splitter.

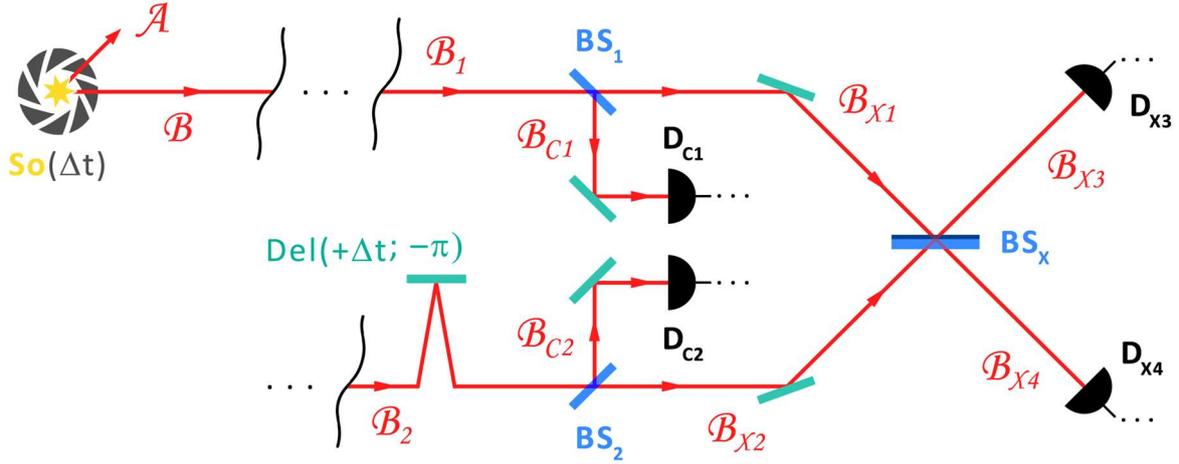

Fig. A8.1. Outline of a construction containing a photon source So(Δt) operating intermittently and a delayer Del; these elements create the conditions for the Hong-Ou-Mandel effect in the F(XOR) gate implemented with the $BS_X$ beam splitter

We will examine two cases: measurement is made in branch A 1) with a 90° angle polarizer, 2) with a 45° angle polarizer.

1) In the case of measuring in branch A with a 90° angle, also taking into account the π phase shift realized by the delayer in branch $B_2$ (using the simplified notation defined at the beginning of the chapter, from Equations (24) and (25) (see chapter 4.5)), the two randomly occurring subcases:

II.a(90°):  $f_{BX1}(90°) \cdot |\psi\rangle_{BX1} + f_{BX2}(90°) \cdot |\psi\rangle_{BX2} = |B_{X1}\rangle$

II.b(90°):  $f_{BX1}(90°) \cdot |\psi\rangle_{BX1} + f_{BX2}(90°) \cdot |\psi\rangle_{BX2} = |B_{X2}\rangle$

Since there are no superposition components in each case, we can interpret them as the photon traveling in either the $B_{X1}$ or $B_{X2}$ branch. After the effect of the delayer Del, four subcases will finally be possible in the detections:

|     | 1st photon emission: $t_0$ | 2nd photon emission: $t_0+\Delta t$ | Delaying effect: detection time difference |
| --- | --- | --- | --- |
| a.) | II.a(90°):  $\|B_{X1}\rangle$ | II.a(90°):  $\|B_{X1}\rangle$ | Δt |
| b.) | II.b(90°):  $\|B_{X2}\rangle$ | II.b(90°):  $\|B_{X2}\rangle$ | Δt |
| c.) | II.a(90°):  $\|B_{X1}\rangle$ | II.b(90°):  $\|B_{X2}\rangle$ | 2·Δt |
| d.) | II.b(90°):  $\|B_{X2}\rangle$ | II.a(90°):  $\|B_{X1}\rangle$ | 0 |

Table A8.1. The possible subcases of detection time differences of photon pairs that passed through the F(XOR) gate when measurements were performed on branch A with a polarizer angle of α=90°. Then, in branch B, in **case II.a(90°)** the state corresponding to the α polarization angle, or in **case II.b(90°)** the state corresponding to α–90° polarization angle, determines the state of the photon entering the F(XOR) gate.

Now we record an important condition: the processing of information from detectors $D_{X3}$ and $D_{X4}$ only considers the coincidences. Generally, by coincidence, we mean the results of detections happening at the same time. Two types of coincidence are possible: identical output coincidence, if two photons exit on the same side of the beam splitter, and non-identical output coincidence, if two photons exit on different sides of the beam splitter. These can be distinguished by using photon number resolving detectors. [22]

Taking into account the coincidences, the cases a.), b.), and c.) in the above table are filtered out. In case d.), one photon arrives at each of the branches $B_{X1}$ and $B_{X2}$ at the beam splitter, entering at the same time. Thus, in this subcase, the Hong-Ou-Mendel effect is formed for the photons exiting the two outputs of the beam splitter (Table A8.2).

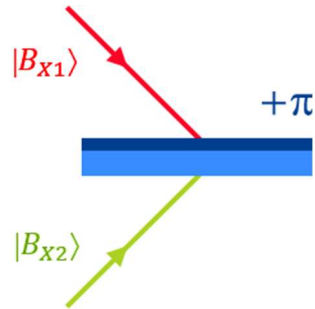

Fig. A8.2. The possible states of two photons entering at the same time the F(XOR) gate implemented with a phase shifter beam splitter, when the measurement was performed with a polarizer with an α=90° angle on branch A.

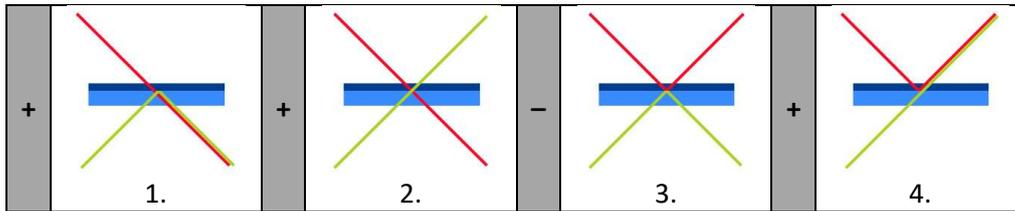

Table A8.2. Inputs defined in Fig. A8.2 and possible outputs of the photons on a phase shifter beam splitter.

The result in simplified form:

$$|B_{X1}\rangle|B_{X2}\rangle \implies f_X(90°) \cdot |\psi\rangle_{BX} = \frac{1}{\sqrt{2}}(-|B_{X3}\rangle|B_{X3}\rangle + |B_{X4}\rangle|B_{X4}\rangle) \qquad (a9)$$

Our conclusion: during the application of a photon source operating intermittently and a delayer on branch B, the cases of two-photon coincidences resulting from measurements on branch A will always – with 100% probability – yield identical output coincidences.

2) In the case of measurements on branch A with a 45° angle, taking into account the π phase shift implemented by the delay on branch $B_2$ (using simplified notation and Equations (22) and (23) (see chapter 4.5)), the two randomly occurring cases are:

II.a(45°): $\quad f_{BX1}(45°) \cdot |\psi\rangle_{BX1} + f_{BX2}(45°) \cdot |\psi\rangle_{BX2} = -\frac{1}{\sqrt{2}} \cdot |B_{X2}\rangle + \frac{1}{\sqrt{2}} \cdot |B_{X1}\rangle$

II.b(45°): $\quad f_{BX}\ (45°) \cdot |\psi\rangle_{BX1} + f_{BX}\ (45°) \cdot |\psi\rangle_{BX2} = \frac{1}{\sqrt{2}} \cdot |B_{X2}\rangle + \frac{1}{\sqrt{2}} \cdot |B_{X1}\rangle$

In both cases, the state of each photon consists of superposition components belonging to the $B_{X1}$ and $B_{X2}$ branches. However, since the $B_{X2}$ branch contains a delayer defined with a time interval of Δt, there is a characteristic difference in time of Δt between the components belonging to this branch and the $B_{X1}$ branch. The consequence of this is that the detection of a photon in such a superposition state occurs with an uncertainty of Δt.

This leads to the following if the photon source emits one photon at intervals of Δt:

– the detection of the first photon occurs at a random time between t and t+Δt;

– the detection of the second photon arriving Δt time later occurs at a random time between t+Δt and t+2·Δt.

This means that in a significant portion of cases, the detectors will not signal a coincidence. Coincidences can occur in a borderline case: when the first photon is detected at time t+Δt within the time interval between t and t+Δt, and the second photon arrives at one of the detectors at time t+Δt within the time interval between t+Δt and t+2·Δt. The relative number of such coincidences (the number of detections relative to the results of the control detectors $D_{C1}$ and $D_{C2}$ in a designated time interval – if we still use the control branches) is presumably much smaller than in case 1) above, where based on Table A8.1 and the mathematical result, there could be coincidences with a 25% probability. This difference in itself means that this type of F(XOR) gate can distinguish between cases – 1) and 2).

We will also examine the borderline case of possible coincidences. Since the photons can randomly enter states II.a(45°) or II.b(45°), for two consecutive photons at the input of the F(XOR) gate, four subcases may occur with equal probability (Table A8.3).

|  | 1st photon | 2nd photon |
|---|---|---|
| a.) | II.a(45°): $-\frac{1}{\sqrt{2}} \cdot \|B_{X2}\rangle + \frac{1}{\sqrt{2}} \cdot \|B_{X1}\rangle$ | II.a(45°): $-\frac{1}{\sqrt{2}} \cdot \|B_{X2}\rangle + \frac{1}{\sqrt{2}} \cdot \|B_{X1}\rangle$ |
| b.) | II.b(45°): $\frac{1}{\sqrt{2}} \cdot \|B_{X2}\rangle + \frac{1}{\sqrt{2}} \cdot \|B_{X1}\rangle$ | II.b(45°): $\frac{1}{\sqrt{2}} \cdot \|B_{X2}\rangle + \frac{1}{\sqrt{2}} \cdot \|B_{X1}\rangle$ |
| c.) | II.a(45°): $-\frac{1}{\sqrt{2}} \cdot \|B_{X2}\rangle + \frac{1}{\sqrt{2}} \cdot \|B_{X1}\rangle$ | II.b(45°): $\frac{1}{\sqrt{2}} \cdot \|B_{X2}\rangle + \frac{1}{\sqrt{2}} \cdot \|B_{X1}\rangle$ |
| d.) | II.b(45°): $\frac{1}{\sqrt{2}} \cdot \|B_{X2}\rangle + \frac{1}{\sqrt{2}} \cdot \|B_{X1}\rangle$ | II.a(45°): $-\frac{1}{\sqrt{2}} \cdot \|B_{X2}\rangle + \frac{1}{\sqrt{2}} \cdot \|B_{X1}\rangle$ |

Table A8.3. The possible states of photon pairs entering the F(XOR) gate when measurements were performed on branch A with a polarizer at an angle of α=45°. Then, in branch B, in **case II.a(45°)** the state corresponding to the α polarization angle, or in **case II.b(45°)** the state corresponding to α–90° polarization angle.

For the examination, let's choose case d). Using the above notation method, we obtain Table A8.4.

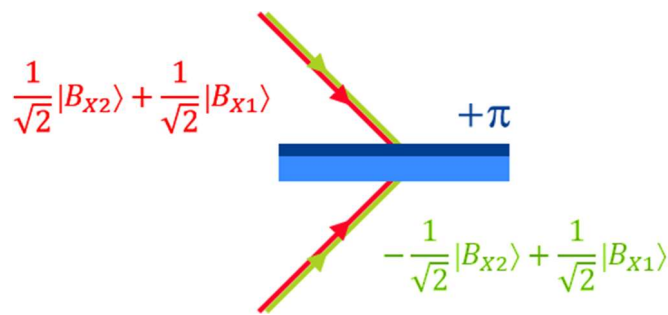

Fig. A8.3. The possible states of two photons entering at the same time the F(XOR) gate implemented with a phase shifter beam splitter, when the measurement was performed with a polarizer with an α=45° angle on branch A.

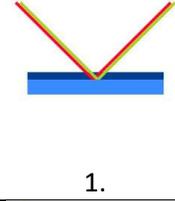
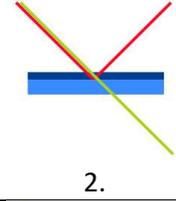
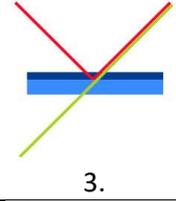
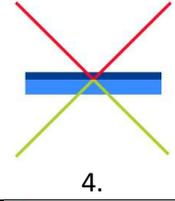
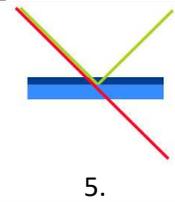
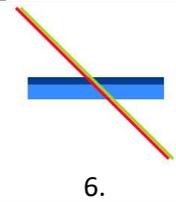
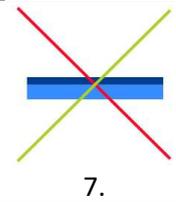
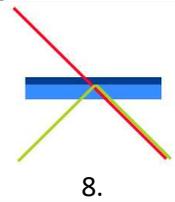
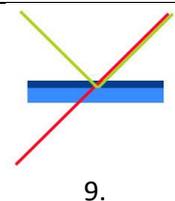
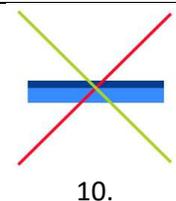
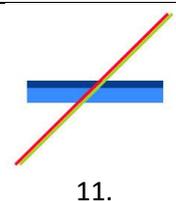
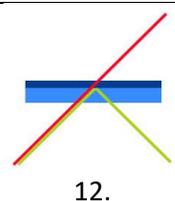
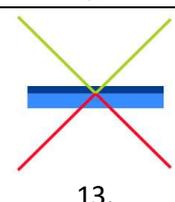
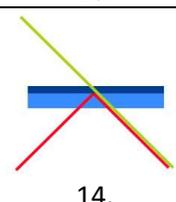
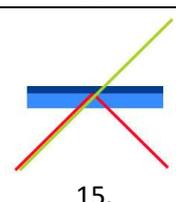
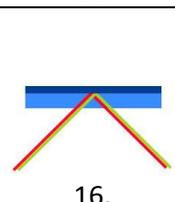

Table A8.4. Inputs defined in Fig. A8.3 and possible outputs of the photons on a phase shifter beam splitter.

Adding up the superpositions, the result in simplified form is:

$$\left(\frac{1}{\sqrt{2}} \cdot |B_{X2}\rangle + \frac{1}{\sqrt{2}} \cdot |B_{X1}\rangle\right)\left(-\frac{1}{\sqrt{2}} \cdot |B_{X2}\rangle + \frac{1}{\sqrt{2}} \cdot |B_{X1}\rangle\right) \Rightarrow$$

$$f_X(45°) \cdot |\psi\rangle_{BX} = -|B_{X3}\rangle|B_{X4}\rangle \qquad (a10)$$

In case d), therefore, different output coincidences occur. It is not necessary to examine the other subcases of Table A8.3 for the following reasons. If at least one subcase results in a proven different output coincidence, in the case of a large photon quantity, when all four subcases appear with equal probability, there will also be non-identical output coincidences in the detection results. In contrast, in case 1), we saw that after filtering out the non-coincidences, the realized coincidences will have identical output coincidences with 100% probability.

So we have found two cases where the F(XOR) gate based on the conditions for inducing the Hong-Ou-Mandel effect led to different detection results. Cases 1.) and 2.) differ on the one hand in the relative degree of coincidences, and on the other hand, they also differ in the proportion of identical and non-identical output coincidences.

Our conclusion: it can be considered proven that there are at least two cases that lead to distinguishable detection results in branch B following two different measurement events in branch A.

As a technical question, it may arise that the above theoretical solution assumes an ideal intermittently operating photon source So(Δt), whose pulses (the photons) are emitted in an infinitely short "time window", since this is the only way to ensure that there is exactly a time difference of Δt between each photons – which is necessary to coincide exactly with the characteristic Δt of the delayer Del. For a realistic source with a "time window," photon emission occurs within Δt ± Δt$_{window}$ time interval. In this case, detectors that exclusively consider coincidences only evaluate photon pairs with a time difference of Δt (and they will be such in the "time windows"), filtering out the rest. Furthermore, coincidences will not only be caused by photon pairs with an infinitely precise time difference Δt, but also in cases where deviations are determined by the quantum mechanical uncertainty principle. However, it is not necessary to detail these aspects for the discussion of the theoretical construction outlined above.

## A9. F(XOR) gate ensuring passage through interference space

The essential part of the F(XOR) gate is formed by a hypothetical interference space (IS) treated as a kind of black box (Fig. A9.1). The B$_{X1}$ and B$_{X2}$ branches are introduced into this space, through which the photons reach the D$_X$ detector. The definition of IS operation is very simple: the detector measures nothing when the phase difference between B$_{X1}$ and B$_{X2}$ imposes total destructive interference.

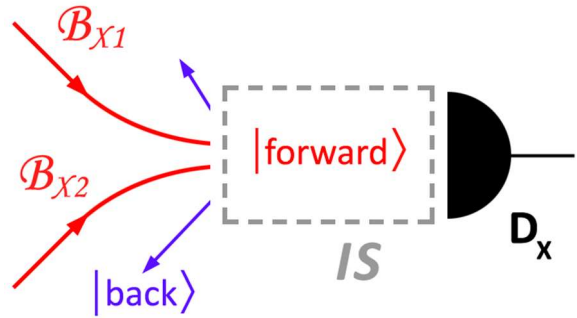

Fig. A9.1. Schematic representation of an F(XOR) gate ensuring passage through a hypothetical interference space

We will examine two cases: 1) measurement is first performed on branch A with a 90° angle, and 2) then with a 45° angle polarizer. We know what components develop on branches B$_{X1}$ and B$_{X2}$ in each case.

1) In the case of measurement on branch A with a 90° angle polarizer, two random subcases occur (based on the simplified notations so far, from equations (24) and (25) (see chapter 4.5)):

II.a(90°): $f_{BX1}(90°) \cdot |\psi\rangle_{BX1} + f_{BX2}(90°) \cdot |\psi\rangle_{BX2} = |B_{X1}\rangle \implies$

$f_X(90°) \cdot |\psi\rangle_{BX} = |\text{forward}\rangle$         (a11)

II.b(90°): $f_{BX1}(90°) \cdot |\psi\rangle_{BX1} + f_{BX2}(90°) \cdot |\psi\rangle_{BX2} = |B_{X2}\rangle \implies$

$f_X(90°) \cdot |\psi\rangle_{BX} = |\text{forward}\rangle$         (a12)

These subcases can also be interpreted as the photon traveling on one or the other branch. In this case, the function of the IS is simply to let the photon pass unchanged to the detector. This means that in both subcases, the photon is detected with a probability of 1. Using the control detectors, the average of the relative probabilities of the two randomly alternating subcases:

$$N_S = \frac{\overline{D_X}}{\overline{D_C}} = \frac{1+1}{2} = 1 \qquad (a13)$$

2) In the case of measurement on branch A with a 45° angle polarizer, the $B_{X1}$ and $B_{X2}$ branches have superposition components that form interference in the IS. It is evident that the passage through the IS occurs according to the probability amplitude resulting from interference. Generally, in the case of destructive interference, the probability of passage will be less than 1, but ensuring the complete state of the photon requires the possibility of backward propagation. This can be done in a controlled manner, but it can also happen uncontrolled (which can be visualized as a kind of "bounce back" for light rays).

Following the measurement on branch A with a 45° angle polarizer, two random subcases occur (using the simplified notation, from equations (22) and (23) (see chapter 4.5)):

II.a(45°): $f_{BX1}(45°) \cdot |\psi\rangle_{BX1} + f_{BX2}(45°) \cdot |\psi\rangle_{BX2} = -\frac{1}{\sqrt{2}} \cdot |B_{X2}\rangle + \frac{1}{\sqrt{2}} \cdot |B_{X1}\rangle \implies$

$$f_X(45°) \cdot |\psi\rangle_{BX} = \frac{1}{\sqrt{2}}\left(\frac{1}{\sqrt{2}} - \frac{1}{\sqrt{2}}\right)|\text{forward}\rangle + \frac{1}{\sqrt{2}}\left(\frac{1}{\sqrt{2}} + \frac{1}{\sqrt{2}}\right)|\text{back}\rangle = |\text{back}\rangle \qquad (a14)$$

II.b(45°): $f_{BX1}(45°) \cdot |\psi\rangle_{BX1} + f_{BX2}(45°) \cdot |\psi\rangle_{BX2} = \frac{1}{\sqrt{2}} \cdot |B_{X2}\rangle + \frac{1}{\sqrt{2}} \cdot |B_{X1}\rangle \implies$

$$f_X(45°) \cdot |\psi\rangle_{BX} = \frac{1}{\sqrt{2}}\left(\frac{1}{\sqrt{2}} + \frac{1}{\sqrt{2}}\right)|\text{forward}\rangle + \frac{1}{\sqrt{2}}\left(\frac{1}{\sqrt{2}} - \frac{1}{\sqrt{2}}\right)|\text{back}\rangle = |\text{forward}\rangle \qquad (a15)$$

These cases are special, since in one case complete destructive interference is realized in IS, while in the other case complete constructive interference. Therefore, in the first case, after the IS, the photons are detected with zero probability, in the second case, the photons are detected with 100% probability. Using the control detectors, the average of the relative probabilities of the two randomly alternating subcases:

$$N_S = \frac{\overline{D_X}}{\overline{D_C}} = \frac{1+0}{2} = \frac{1}{2} \qquad (a16)$$

Our conclusion: an F(XOR) gate based on the principle of interference space can be determined in terms of quantum computing, which leads to different detection results in cases II(90°) and II(45°). This means that two different measurement events on branch A can lead to distinguishable information on branch B.

We need to make a fundamental observation. In the two cases, the photons behave differently within the IS. If we interpreted case 1) as the photons arriving on one or the other branch, then they pass through the IS without change. There is no reason to consider the possibility of backward propagation. However, in the first subcase of case 2.), it is necessary to take this into account, since in the IS the two superposition components form a complete destructive interference. These two possible behaviors of the photons distinguish this solution from the solution where the two branches run together in a simple beam splitter. There, only one type of behavior was allowed for the photons: the wave components were always distributed symmetrically on both sides of the beam splitter. As we summarized in chapter 4.3, the consequence of this was that the beam splitter could not distinguish between the different cases.

The mathematical description above is valid only when the two branches meet each other under 0° (parallel). Only in this case, the probability amplitudes given by Equations (a14) and (a15) can be ensured in the cross section of the IS. In the case of angles other than this, interferences occur in the cross-section, which degrade and invalidate those determined by the formulas, the average detection values of the two cases will be equal. The technical solution of this is certainly not a trivial task. It is possible that further development of solutions where two inputs are connected into one output, as in the case of [23], could lead to similar results.

We cannot rule out the possibility that the physical feasibility of the quantum information principle underlying the calculations may eventually be constrained by some quantum mechanical principle, which might imply that the two types of photon behavior described above are not possible. However, it is also conceivable that several implementations are possible, although these are likely to be non-trivial. (It is possible that it can be done by combining optical modes.) Nevertheless, purely from a quantum information perspective, the operational principle of the hypothetical IS is quite simple.